\documentclass[letterpaper]{article} % DO NOT CHANGE THIS
\usepackage{aaai2026}  % DO NOT CHANGE THIS
\usepackage{times}  % DO NOT CHANGE THIS
\usepackage{helvet}  % DO NOT CHANGE THIS
\usepackage{courier}  % DO NOT CHANGE THIS
\usepackage[hyphens]{url}  % DO NOT CHANGE THIS
\usepackage{graphicx} % DO NOT CHANGE THIS
\usepackage{natbib}  % DO NOT CHANGE THIS AND DO NOT ADD ANY OPTIONS TO IT
\usepackage{caption} % DO NOT CHANGE THIS AND DO NOT ADD ANY OPTIONS TO IT
\usepackage{algorithm}
\usepackage[noend]{algpseudocode}
\usepackage{amsmath}
\usepackage{float}
\usepackage{amssymb}
\usepackage{booktabs}\usepackage{graphicx}
\usepackage{subcaption}
\usepackage{comment}
\usepackage{multirow}
\usepackage{tabularx}
\usepackage{newfloat}
\usepackage{listings}
\DeclareCaptionStyle{ruled}{labelfont=normalfont,labelsep=colon,strut=off} % DO NOT CHANGE THIS
\floatstyle{ruled}
\newfloat{listing}{tb}{lst}{}
\floatname{listing}{Listing}
\title{Enhancing PIBT via Multi-Action Operations}
\author{
   Egor Yukhnevich$^{1}$, Anton Andreychuk$^{2}$
}
\affiliations{
    \textsuperscript{\rm 1}HSE University\\
    \textsuperscript{\rm 2}Cognitive AI Systems Lab\\
    Moscow, Russia\\
    evyukhnevich@edu.hse.ru, andreychuk@cogailab.com
}

\usepackage{bibentry}
\begin{document}

\maketitle

\begin{abstract}
PIBT is a rule-based Multi-Agent Path Finding (MAPF) solver, widely used as a low-level planner or action sampler in many state-of-the-art approaches. Its primary advantage lies in its exceptional speed, enabling action selection for thousands of agents within milliseconds by considering only the immediate next timestep. However, this short-horizon design leads to poor performance in scenarios where agents have orientation and must perform time-consuming rotation actions. In this work, we present an enhanced version of PIBT that addresses this limitation by incorporating multi-action operations. We detail the modifications introduced to improve PIBT's performance while preserving its hallmark efficiency. Furthermore, we demonstrate how our method, when combined with graph-guidance technique and large neighborhood search optimization, achieves state-of-the-art performance in the online LMAPF-T setting.
\end{abstract}

\begin{links}
    \link{Project page}{https://sites.google.com/view/epibt}
    %\link{Extended version}{https://aaai.org/example}
\end{links}

\section{Introduction}

\begin{figure}[t]
	\includegraphics[width=\linewidth]{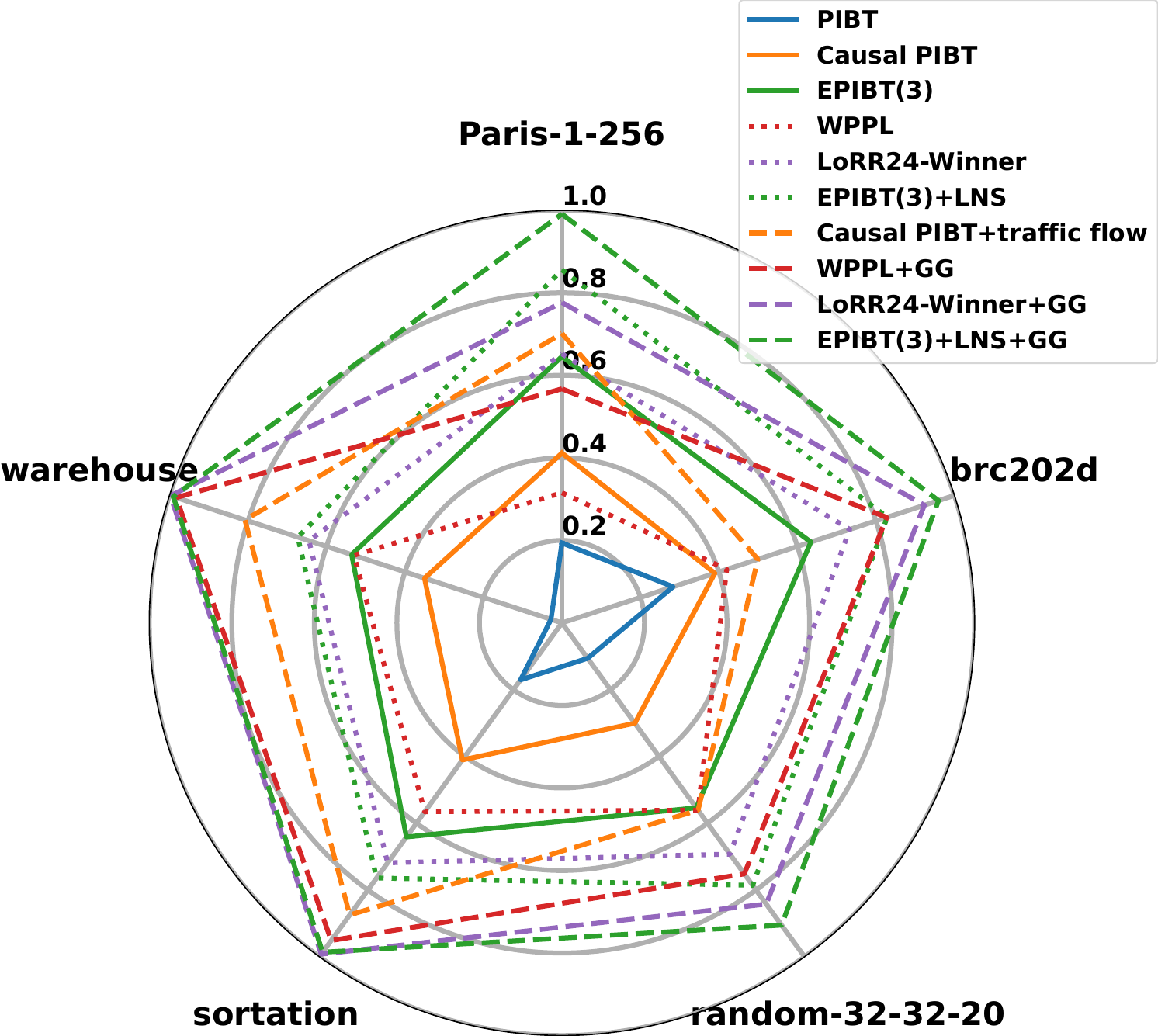}
	\caption{Spider plot demonstrating the relative performance of the evaluated approaches. Solid lines represent PIBT-like approaches without additional components, dotted lines indicate methods utilizing LNS, and dashed lines denote approaches that also incorporate GG. Related solvers are indicated by the same color.}
    \label{fig:spider_plot}
\end{figure}

Multi-agent Pathfinding (MAPF) is a well-known and extensively studied problem in which a group of agents, starting from their initial locations, must reach designated goal locations while avoiding collisions. Numerous variations of this problem exist~\cite{stern2019multi}. The search for an optimal solution to the classical MAPF problem is known to be NP-hard~\cite{geft2022refined}. Consequently, existing MAPF solvers that guarantee finding an optimal solution -- such as CBS~\cite{sharon2015conflict} and its variants~\cite{li2019improved,boyarski2015icbs}, BCP~\cite{lam2022branch}, ICTS~\cite{sharon2013increasing}, and others -- face significant challenges with runtime and scalability as the number/density of agents increases. To address scalability problems, suboptimal variants~\cite{barer2014suboptimal,huang2021learning,li2021eecbs} and anytime approaches, such as MAPF-LNS~\cite{li2022mapf,huang2022anytime} and LaCAM*~\cite{okumura2023lacam,okumura2024engineering}, have been developed. Anytime solvers often employ extremely fast rule-based techniques, such as PIBT~\cite{okumura2022priority} or Push-and-Rotate~\cite{de2014push}, to quickly generate initial solutions, which are then iteratively improved within a given time budget.

However, the aforementioned solvers are tailored for the classical MAPF scenario, in which all goal locations are known in advance. Some algorithms explicitly leverage this assumption; for instance, LaCAM employs depth-first search to accelerate the search process. In contrast, there exists a variant of the MAPF problem where agents are assigned new goal locations each time they reach their current goal. This variant is commonly referred to as Lifelong MAPF (LMAPF). The absence of information about future goals at the outset makes most traditional MAPF solvers unsuitable for these scenarios, necessitating specific adaptations or entirely new approaches. One of the most common and effective strategies for addressing the challenge of continuously updating goals is the use of windowed search-planning only for a limited number of upcoming timesteps rather than the entire planning horizon. Consequently, when agents reach their current goals and receive new ones, the algorithm can be relaunched to generate the next segment of the plan, explicitly incorporating the new goals. The concept of windowed search for MAPF/LMAPF problems was first introduced in \cite{silver2005cooperative} and has since been successfully adopted in approaches such as RHCR~\cite{li2021lifelong}, WPPL~\cite{jiang2024scaling}, WinC-MAPF~\cite{veerapaneni2025windowed} among others.

Another advantage of windowed-based solvers is their ability to quickly return the next actions, making them suitable for the online LMAPF problem. In this setting, the algorithm operates concurrently with real agents as they execute the planned actions. Consequently, the algorithm must provide the next set of actions within a strict time frame, often as short as one second. This requirement is imposed by real-world applications, such as autonomous warehouses \cite{dhaliwal2020rise}, logistics \cite{ferreira2023systematic}, and others.

Another practical constraint, frequently encountered in real-world scenarios but often overlooked in MAPF/LMAPF literature, is the presence of agent orientations and the associated time required to change them. While most approaches can be adapted to action models that include rotations with relatively straightforward modifications, some methods require significant changes to maintain their performance and efficiency when applied to such models.

In this work, we present a modification of the rule-based iterative solver, Priority Inheritance with Back Tracking (PIBT)~\cite{okumura2022priority}, which is known for its excellent scalability. However, due to its design, PIBT struggles when applied to agents with a rotation action model, as agents may require more than one timestep to vacate an occupied cell if they must rotate first. The core idea of our proposed modification is the introduction of multi-action operations that combine all possible actions -- rotations, moves, and waits. We also demonstrate that this modification can be beneficial not only for the rotation action model but also for the classic omnidirectional action model. Furthermore, our experimental evaluation shows that combining the modified PIBT with additional techniques, such as graph guidance and large neighborhood search, enables it to outperform existing state-of-the-art approaches on the considered problem statement: online LMAPF-T, i.e., Lifelong MAPF with a limited time budget and a rotation action model.

\section{Problem Statement}
The environment is represented as a 4-connected grid of cells, where each cell is identified by its coordinates $(i, j)$. Each cell can be either traversable or blocked (representing obstacles). The set of all traversable cells is denoted by $V \subseteq \mathbb{Z}^2$, and the set of edges $E$ consists of pairs of adjacent traversable cells, i.e., $E = \{ ((i, j), (i', j')) \mid |i - i'| + |j - j'| = 1,\, (i, j) \in V,\, (i', j') \in V \}$.

We are given $n$ agents, each with a unique start location and orientation. The set of start locations is $\{(i_1^0, j_1^0), \ldots, (i_n^0, j_n^0)\}$, where $(i_k^0, j_k^0) \in V$ for $k = 1, \ldots, n$. Each agent also has an initial orientation $o_k^0 \in \{\text{north}, \text{east}, \text{south}, \text{west}\}$. The state of each agent at time $t$ is thus defined by both its location and orientation, i.e., $s^t_k = ((i_k^t, j_k^t), o_k^t)$.

An external task assignment module provides each agent with a current goal location $g_k = (i_{g_k}, j_{g_k}) \in V$. Whenever an agent reaches its current goal, it is immediately assigned a new goal location by the task assignment module. We assume that only the current goal location for each agent is known at any given time; future goals are revealed dynamically as agents reach their current goals.

We consider the \emph{rotation action model}, in which each agent can perform one of four actions at each discrete timestep: $F$ -- move forward, $R$ -- rotate $90^\circ$ clockwise, $C$ -- rotate $90^\circ$ counterclockwise, $W$ -- wait in place. All actions have unit duration. %Thus, at each timestep, an agent may move forward to a traversable adjacent cell, rotate $90^\circ$ in either direction, or wait. To reverse direction (rotate $180^\circ$), an agent must perform two consecutive rotation actions. 

A collision occurs if, at any timestep, either (i) two agents occupy the same cell ($(i_k^t, j_k^t) = (i_l^t, j_l^t)$ for $k \neq l$), or (ii) two agents traverse the same edge in opposite directions simultaneously ($( (i_k^{t-1}, j_k^{t-1}), (i_k^t, j_k^t) ) = ( (i_l^t, j_l^t), (i_l^{t-1}, j_l^{t-1}) )$ for $k \neq l$).

The objective is to maximize the total number of goals reached by all agents within a fixed timestep horizon $T$, while avoiding collisions. This objective, commonly referred to as \emph{throughput}, is defined as the average number of goals reached per timestep.

Additionally, we impose a strict time budget for the solver to compute the next actions for all agents at each timestep. This constraint is motivated by real-world applications, where the planner must react promptly to changes and operate in an online regime. In our experiments, this time limit is set to 1 second per timestep. If the solver exceeds this limit, all agents are delayed by one timestep for each additional second spent. The time budget is not cumulative: even if the solver uses less than the allotted time in one step, it cannot carry over the unused time to subsequent steps.

The problem statement considered in this work is identical to that of the League of Robot Runners\footnote{https://www.leagueofrobotrunners.org/} (LoRR)~\cite{chan2024league} -- a competition sponsored by Amazon Robotics, designed to bridge the gap between fundamental research and industrial applications, including warehouse logistics, transportation, and advanced manufacturing. Notably, the most recent edition of the competition introduced new challenges, such as dynamic task assignment and sequences of tasks. In this work, we focus exclusively on efficient agent routing, assuming that task assignment is handled by an external module beyond our control.

\section{Priority Inheritance with Backtracking}

Before detailing the modifications introduced in the PIBT approach, it is useful to briefly describe its core concept. PIBT is a windowed solver with a single-step window, meaning that it provides collision-free actions for all agents for only one timestep. To achieve this, PIBT employs a prioritized approach: each agent is assigned a priority based on its distance to its goal, and agents plan their actions sequentially, avoiding collisions with higher-priority agents and selecting the most beneficial action -- typically, the one that reduces their distance to the goal.

A key feature of PIBT is the use of priority inheritance with backtracking. If an agent $k$ encounters a collision with a higher-priority agent $l$, it inherits the priority $p_l$ and attempts to vacate the occupied cell, potentially pushing any other agents with lower priority than $p_l$. If such a push is unsuccessful, the agent receives a signal via backtracking and attempts to select an alternative action.

It was shown in~\cite{okumura2022priority} that all agents will eventually reach their goal locations. However, simultaneous occupation of goal locations by all agents is not guaranteed. Additionally, a condition on the graph structure must be satisfied: for every pair of adjacent nodes, there must exist a simple cycle $C$ containing both nodes, with $|C| \geq 3$. This ensures that the agent with the highest priority can always push other agents from the desired location, guaranteeing that it will reach its goal within a bounded number of steps. After reaching its goal, its priority is decreased, enabling the next agent to reach its goal in a similarly bounded number of steps.

The original PIBT approach assumes that, at each timestep, at least one agent with the highest priority can move closer to its goal, as it can always vacate the desired location. However, under the rotation action model, this assumption does not hold: the agent occupying the desired location may first need to rotate before it can vacate the cell. This issue can be partially addressed by the Causal PIBT approach~\cite{okumura2021time}, which considers event-based causal dependencies between agents’ actions. Causal PIBT was initially developed for a stochastic setting, where each action may be delayed with some probability; in the present context, rotation actions can be interpreted as deterministic delays.

Another way to address delays caused by rotation actions is to consider multiple timesteps, rather than just one. The idea of planning over multiple steps was explored in~\cite{okumura2019winpibt}, where the winPIBT approach was introduced. Instead of planning only the next action, agents plan a path for the next $w$ steps. This algorithm is designed for classical MAPF and aims to resolve blockages caused by agents that have already reached their goal locations.

In contrast to winPIBT, which considers relatively long windows (up to 30 steps in the original paper), the approach proposed here uses a much shorter window (up to 5 steps). Rather than launching a path-planning algorithm, our method considers multi-action operations of limited length.

\section{Enhanced PIBT}

In this section, we describe the modifications made to the original PIBT approach to efficiently support the rotation action model. In addition to introducing multi-action operations, we allow agents to be revisited during a single step; that is, an agent may change its chosen action even if a valid collision-free action has already been found. We also employ inheritance of operations, enabling the reuse of actions selected by agents in the previous step. The approach that integrates all these modifications is referred to as Enhanced PIBT, or simply EPIBT.

\begin{figure}[t]
    \centering
    \includegraphics[width=\linewidth]{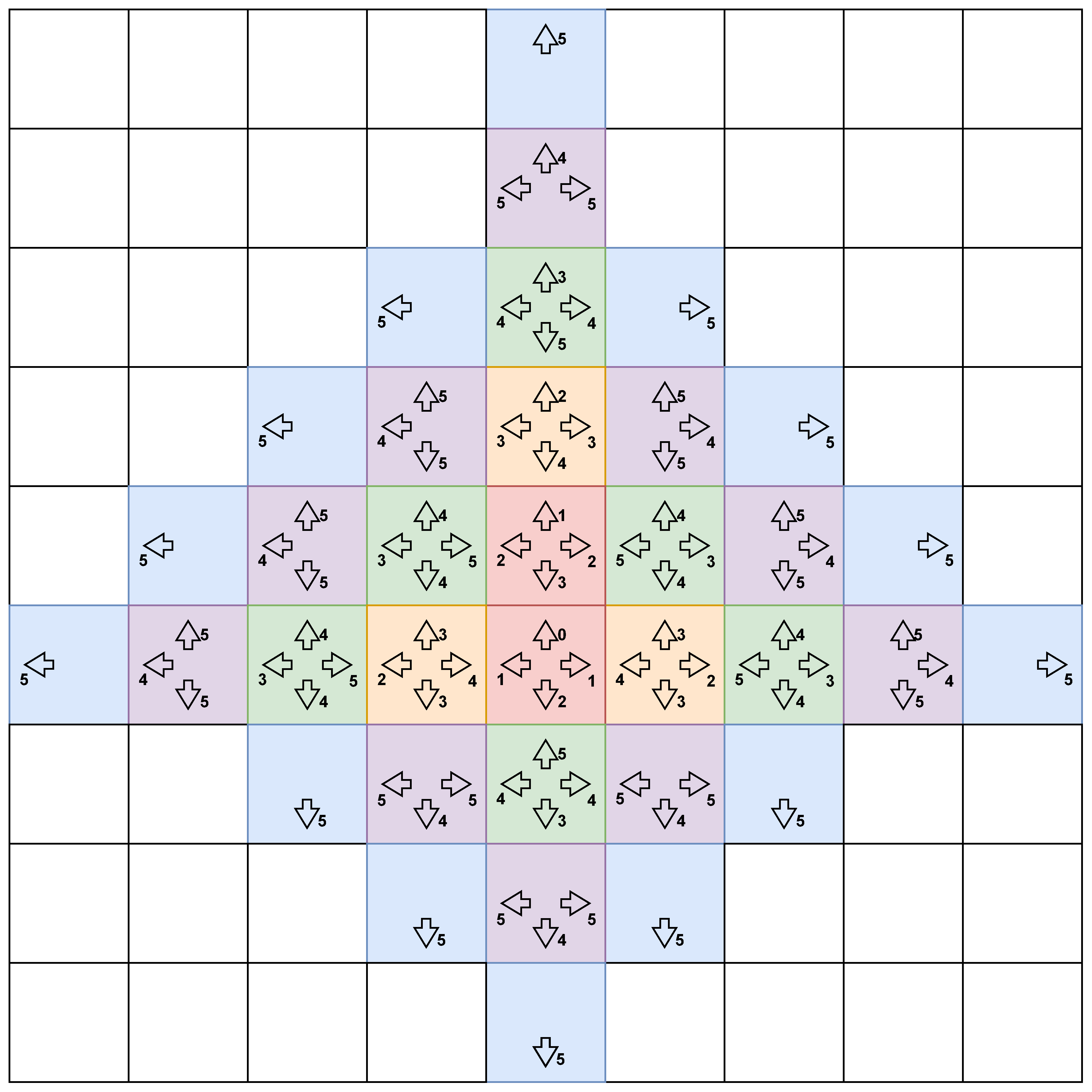}
    \caption{Cells and states reachable with different operation length. Different cell colors indicate the minimum required operation length to reach the corresponding cell. Arrows and numbers near them indicate the actual state and number of actions required to reach it.}
    \label{fig:epibt_cells}
\end{figure}

\begin{table}[t]
    \centering
    \begin{tabular}{|c|ccccc|}
        \hline
        Length & 1 & 2 & 3 & 4 & 5 \\
        \hline
        Cells & 2 & 5 & 11 & 21 & 35\\
        States & 4 & 10 & 23 & 48 & 88\\
        Cells sequences & 2 & 6 & 17 & 48 & 136\\
        \hline
    \end{tabular}
    \caption{Number of possible reachable cells, reachable states and unique sequences of occupied cells depending on operation length.}
    \label{tab:epibt_cells}
\end{table}

\subsection{Multi-action Operations}
A central aspect of EPIBT is the concept of operations. Each operation consists of a sequence of actions performed by an agent. Figure~\ref{fig:epibt_cells} illustrates all cells and states that are reachable with operations of different lengths. With an operation length of 1, as in regular PIBT, only two cells (marked in red) are reachable. This limited reachability significantly restricts the ability to efficiently resolve collisions between agents.

Although the maximum number of possible action combinations is $4^{\text{length}}$, the actual number of operations that must be considered is much smaller, as many operations can be discarded. First, operations containing multiple redundant rotation actions can be eliminated. Additionally, it is unnecessary to consider operations that end with rotations, since rotations do not change the agent's location. Instead, multiple operations such as $FFW$\footnote{The notation $FFW$ means that the agent consequently executes actions move forward, move forward and wait.}, $FFR$, and $FFC$ can be merged into a single $FFW$ operation, with the $h$-value (distance to the goal) of the successor state calculated based on the best heading among the reachable ones. However, the number of operations still exceeds the number of possible reachable states. Due to the presence of other agents in the workspace, it is necessary to consider the time dimension and include wait actions in operations, allowing an agent to reach a state at a later time if needed to avoid collisions. The actual number of operations that must be considered corresponds to the number of possible sequences of cells that the agent occupies while executing the operation. Table~\ref{tab:epibt_cells} shows how many different cells and states can be reached depending on the operation length, as well as the number of unique cell sequences. For clarity, cells refer to locations with $(i, j)$ coordinates, states additionally include orientation (i.e., are defined by the tuple $((i, j), o)$), while cell sequences are sequences of locations with length corresponding to the operation length -- e.g., $\{(i_1, j_1), (i_2, j_2), (i_3, j_3)\}$ for length 3.

Another important aspect of operations is the order in which they are considered. In regular PIBT, actions are prioritized based on the distance to the goal. In EPIBT, this approach must be extended, as multiple states may have the same $h$-value. To address this, we implemented a tie-breaking mechanism that favors forward movement over rotation, and rotation over wait. Empirically, this mechanism produced the best results.

\subsection{Revisiting Agents}
Multi-action operations provide agents with a significantly greater set of options. While this is beneficial for performance, it also introduces new challenges. The first challenge is that considering multiple timesteps can lead to multi-agent collisions, i.e., situations where a chosen operation results in a collision with more than one agent. In such cases, recursively invoking the operation selection for each colliding agent leads to ambiguity, as these agents may mutually affect each other and all inherit the same priority due to priority inheritance. To address this, the proposed approach does not consider operations that would result in collisions with more than one agent.

The second challenge arising from the increased variety of options is the diversity of possible collisions between agents. In the original PIBT, each agent may be visited only once per timestep, and once its action is chosen, it cannot be changed. This restriction significantly limits the options available to agents. Moreover, if some agents fail to find collision-free operations, they will remain stationary for at least the next timestep. To overcome this limitation, we allow agents to be revisited, i.e., to reselect their operations in order to better resolve collisions and to find collision-free actions for at least some agents that have already been visited but failed to find valid operations. However, in high-density scenarios, unlimited revisiting can lead to a substantial increase in the algorithm's runtime. To mitigate this, we introduce a limit $L$ on the number of times each agent can be revisited during a single timestep.

\begin{algorithm}[b!]
\caption{EPIBT - Main Loop}
\label{alg:epibt_main}
\begin{algorithmic}[1]

\State \textbf{Input:} \( G \), \(\{s_1, \dots, s_n\}\), \(\{g_1, \dots, g_n\}\), $\{a'_1, \dots,a'_n\}$, $L$
%\State \textbf{Output:} selected actions \(\{a_1, \ldots, a_n\}\)

\State \(a_k \leftarrow a'_k\) for $k = 1, \dots, n$
\State \(visited_k, hit_k \leftarrow \{0\}\) for $k = 1, \dots, n$
\State \(P \leftarrow \text{getPath}(s_k, a_k) \text{ for } k = 1, \dots, n\)

\State \(p_k \leftarrow dist(s_k, g_k)\) for each agent \(k = 1, \dots, n\)
\State \(Agents \leftarrow \text{sort }\{1, \dots, n\}\) by priorities \(p_k\)
\For{\(k \in Agents\)}
    \If{\(visited_k \neq 0\)} \textbf{continue} \EndIf
    \State $P \leftarrow P\setminus \text{getPath}(s_k, a_k)$
    \If{\(\text{EPIBT}(k, p_k) = failed\)}
        \State $P \leftarrow P \cup \text{getPath}(s_k, a_k')$
    \EndIf
\EndFor
\State \( \textbf{return } \{a_1, \dots, a_n\}\)

\end{algorithmic}
\end{algorithm}

\subsection{Inheritance of Operations}
Another important feature implemented in EPIBT is the inheritance of operations. Like regular PIBT, EPIBT is executed at every timestep. When constructing operations from scratch, we assume that all agents will occupy their current locations for the next few steps. This assumption is necessary to detect potential collisions at any considered timestep and to allow EPIBT to recursively attempt to resolve them. However, in the previous step, we may have already constructed collision-free operations for multiple timesteps, of which only the first actions have been executed. Therefore, we can reuse the operations built in the previous step by removing the first action (as it has already been executed) and appending a wait action at the end. This preserves the operation length and ensures that these operations remain collision-free.

It is important to note that operation inheritance is used only for the initialization of operations. Agents are still free to select any operation, and there is no requirement that the newly chosen operation must match the previous one in its initial actions. This initialization helps to reduce the number of collisions among agents. Furthermore, if some agents fail to find collision-free operations in the current step, they continue to execute the operations found in the previous step, rather than simply waiting in place.

\begin{algorithm}[t!]
\caption{EPIBT - Operation Selection Procedure}
\label{alg:epibt_procedure}
\begin{algorithmic}[1]
\State \textbf{Input:} \( \text{agent } k\text{, priority } p\) 
    \State \(OP \leftarrow \) sort \(op \in Operations\) by weights \(w_{op}\)
    \State \(visited_k \leftarrow visited_k + 1\); \(hit_k \leftarrow 1\)
    \For{\(op \in OP\)}
        \If{getPath$(s_k, op) \not\subset G$} \textbf{continue} \EndIf
        
        \If{\(\text{getUsed}(s_k, op, P) = \emptyset\)}
            \State $a_k\leftarrow op$; \(hit_k \leftarrow 0\)
            \State \(P \leftarrow P \cup \text{getPath}(s_k, op)\)
            \State \(\textbf{return } success\)
        \EndIf

        \If{$|\text{getUsed}(s_k, op, P)| > 1$} \textbf{continue} \EndIf
        \State \(l \leftarrow \text{getUsed}(s_k, op, P)\)
        \If{$hit_l = 1$ \textbf{or} $visited_l \geq L$ \textbf{or} $p_l \leq p$} 
        \State \(\textbf{continue} \)
        \EndIf

        \State $P \leftarrow P \setminus \text{getPath}(s_l, a_l) \cup \text{getPath}(s_k, op)$
        \State $a_k\leftarrow op$

        \If{EPIBT($l, p$)$ = success$}
            \State \(hit_k \leftarrow 0\); \textbf{return} $success$
        \EndIf

        \State $P \leftarrow P \setminus \text{getPath}(s_k, op) \cup \text{getPath}(s_l, a_l)$
    \EndFor
    \State $a_k \leftarrow a_k'$; \(hit_k \leftarrow 0\)
    \State \textbf{return} $failed$

\end{algorithmic}
\end{algorithm}

\subsection{Pseudocode}

The pseudocode for EPIBT is shown in Algorithm~\ref{alg:epibt_main} (Main Loop) and Algorithm~\ref{alg:epibt_procedure} (Operation Selection Procedure), which together capture all EPIBT enhancements. The main loop takes as input the grid-graph $G$, current agent states $\{s_1, \ldots, s_n\}$, goal locations $\{g_1, \ldots, g_n\}$, inherited operations $\{a'_1, \ldots, a'_n\}$, and revisit limit $L$.

First, the chosen operations $\{a_1, \ldots, a_n\}$ and agent paths $P$ (the states visited while executing actions) are initialized (lines 2--4). Agents are then prioritized by their current distance to goal (lines 5--6). The main loop iterates over unvisited agents, allowing each to select an operation. If no collision-free operation is found, the agent reverts to its inherited operation (lines 8--12).

Algorithm~\ref{alg:epibt_procedure} details the operation selection. Each agent considers operations in order of
$w_{op} = h(s_k, op, g_k) \cdot \alpha + \beta_{op}$, where $h(s_k, op, g_k)$ is a function returning the distance to goal $g_k$ after performing operation $op$ from state $s_k$, and $\alpha$ and $\beta_{op}$ are weighting coefficients. In our implementation, $\alpha$ is set to a high value, making $\beta_{op}$ primarily a tie-breaker when two states have the same $h$-value. The $\beta$ values are chosen such that movement actions are most preferred, followed by rotations, and finally wait actions (line 2). The agent is then marked as visited for this call, and a flag is set to prevent multiple visits within the same recursion branch (line 3).

For each viable operation, \texttt{getPath}$(s_k, op)$ returns the sequence of states. Operations leading to obstacles or outside the grid are skipped (line 5). Valid operations are checked for agent collisions using \texttt{getUsed}$(s_k, op, P)$, which returns IDs of agents that would collide with $k$. Collision-free operations are adopted (lines 6--9). Operations causing collisions with two or more agents are skipped (line 10). For a single-agent collision ($l$), the operation is skipped if $l$ was already visited in this recursion branch, exceeded its revisit limit, or has higher priority (lines 12--13). Otherwise, the algorithm attempts to rebuild $l$'s actions, considering $k$'s operation. If successful, $k$ adopts the operation and returns success (lines 15--17); otherwise, $P$ is reverted (line 18) and the next operation is tried. If all options fail, the agent is assigned its inherited operation $a'_k$ (line 19), and failure is returned (line 20).

\section{Large Neighborhood Search}

EPIBT produces valid, collision-free actions within milliseconds, but these are generated in a prioritized, order-dependent manner. To further improve solution quality and utilize the remaining computational time, we incorporate a Large Neighborhood Search (LNS) optimization. Similar approach, combining windowed PIBT with LNS, have been used in WPPL~\cite{jiang2024scaling}, and LNS-based methods have also been proposed for classical MAPF~\cite{li2021anytime,li2022mapf}. Our goal is not to advance LNS itself, but to show that combining EPIBT with LNS can substantially boost performance, even surpassing state-of-the-art solvers for online LMAPF-T.

At each LNS iteration, a random agent $k$ is selected, its operation and path are removed, and a the operation selection procedure is performed. To enable diverse solutions across runs, the agent $k$ gets the highest priority, allowing it to override the others. If a new solution is found, we accept it if it improves the LNS metric, defined as the sum of $w_{op} \cdot p_k$ for all agents, where $p_k$ reflects proximity to the goal. This metric encourages agents to complete tasks quickly and move on to new ones. The optimization process continues until the time budget is exhausted.

\section{Theoretical Analysis}
In this section, we demonstrate and prove that none of the proposed modifications made in EPIBT violate the major properties of the original PIBT algorithm. Note that the properties of PIBT are valid only when all adjacent nodes in the graph have a simple cycle $C$ such that $|C| \geq 3$. The EPIBT approach requires this assumption as well.

First, we prove the following lemma, similar to Lemma 1 in \cite{okumura2022priority}:

\textbf{Lemma 1}. Let $k$ denote the agent with the highest priority at timestep $t$, and let $(i,j)$ be the location nearest to $g_k$ among the neighbors of $s_k$. In the worst case, agent $k$ will reach location $(i,j)$ within $t+3$ timesteps.

For simplicity, we consider only operations of length 3. The logic of the proof can be adapted to the higher lengths of operations by adding wait actions to the end of operations.

\textbf{Proof}. Consider the very first timestep or the case when no operation inheritance enhancement is applied. All agents have the initial operation WWW, which is guaranteed to be collision-free. Agent $k$ with the highest priority chooses its operation first, while all other agents still have predefined operations WWW. 

The optimal operation might be FFF, FCF, or similar operations that involve more than one movement action, potentially causing collisions with multiple agents. Following the logic of Algorithm~\ref{alg:epibt_procedure} (line 10), such operations are skipped. However, since location $(i,j)$ is an adjacent cell near the current state $s_k$, it can be reached by one of the following operations: WWF, CWF, RWF, or RRF. These four operations cover all adjacent locations, one of which is guaranteed to be closer to $g_k$ than state $s_k$. Moreover, all these operations have only one movement action performed as the final action, which means that: (i) they may result in collision with at most one agent, and (ii) the colliding agent has two extra timesteps to change orientation before freeing the location.

If the desired location $(i,j)$ is free, agent $k$ can obtain the corresponding operation (lines 6-9 of Algorithm~\ref{alg:epibt_procedure}) and will not be forced to change the chosen operation since it has the highest priority. Otherwise, a recursive call of the EPIBT procedure for agent $\ell$ (currently occupying location $(i,j)$) is made (line 16 of Algorithm~\ref{alg:epibt_procedure}). 

The only restriction for agent $\ell$ is that it cannot collide with agent $k$. Since all locations have at least two traversable adjacent cells (due to our assumption about simple cycles with length $\geq 3$), agent $\ell$ can move to some location $(i',j')$ that differs from both location $(i,j)$ and the location of state $s_k$. Even if agent $\ell$ has a wrong orientation and cannot immediately move to location $(i',j')$, it has two extra timesteps to change orientation before movement.

If location $(i',j')$ is also occupied by some agent $m$, it can be pushed by agent $\ell$ out of this location, considering priority inheritance. Recursively applying the logic of procedure EPIBT, agent $m$ can free the occupied location within at most 3 timesteps, and agent $\ell$ can wait in its current location if necessary (by considering one of the four operations mentioned above). Thus, regardless of the number of affected agents, each can move out of the occupied location within 3 timesteps in the worst case. Therefore, agent $k$ with the highest priority will be able to occupy a location closer to $g_k$ than the current state $s_k$ within at most 3 timesteps.

Applying operation inheritance leads to changes in the initial operations that agents have. However, the inherited operations remain collision-free, meaning that even in the worst case, the agent with the highest priority can choose an operation that replicates the inherited operation and changes only the last action. Such operation may collide with at most one agent. If such a collision exists, it occurs at the end of the operation and involves a single agent $\ell$. Thus, it can be successfully resolved by pushing agent $\ell$ away. The same logic applies to the operation choices of the remaining agents, which can be recursively visited while freeing the cell for agent $k$. In the worst case, they may all replicate the inherited operations, changing only the last action. Therefore, operation inheritance does not violate the property that agent $k$ with the highest priority needs at most 3 timesteps to get closer to its goal location $g_k$, even in the worst case. $\square$

Second, we perform an analysis of the time complexity of EPIBT.

\textbf{Proposition 1}. The time complexity of EPIBT in one timestep is $O(n (log\space n + |OP| \cdot log |OP| + L\cdot |OP|\cdot op\_len))$, where $n$ stands for the number of agents, $|OP|$ -- the number of operations considered, $op\_len$ -- operations length, and $L$ -- revisit limit.

\textbf{Proof}. Here we assume (and made it in our implementation) that all distance matrices are precalculated. Thus, the procedure of agents sorting in the main loop of EPIBT costs $O(n\cdot log\space n)$, and the sorting of operations costs $O(|OP| \cdot log |OP|)$. Procedures related to collision checks and reservation updates depend on the operation length, and thus cost $\Theta(op\_len)$. Each operation selection procedure call examines at most $|OP|$ operations that result in $O(|OP| \cdot log |OP| + |OP|\cdot op\_len)$. The sorting of operations is performed once for each agent, while operation selection is performed up to $L$ times, since each agent is visited at most $L$ times per timestep. Therefore, the per-timestep complexity is $O(n (log\space n + |OP| \cdot log |OP| + L\cdot |OP|\cdot op\_len))$. $\square$

Lastly, in \cite{okumura2022priority}, a theorem was proved guaranteeing that all agents will reach their goal locations in a finite number of steps. However, it is not guaranteed that all goal locations will be occupied by the corresponding agents simultaneously, which is required by the classical one-shot MAPF problem. This property is achieved by modifying priorities: each agent that reaches its goal location receives a lower priority than other agents that have not yet reached their goals. Thus, the agent with the highest priority is always one that has not yet reached its goal location, until no such agent exists.

In case of LMAPF neither simultaneous nor eventual goal reaching is required. Nevertheless, EPIBT can utilize the same priority mechanism presented in the original PIBT approach and obtain the same property if required.

\begin{figure}[t!]
	\includegraphics[width=\linewidth]{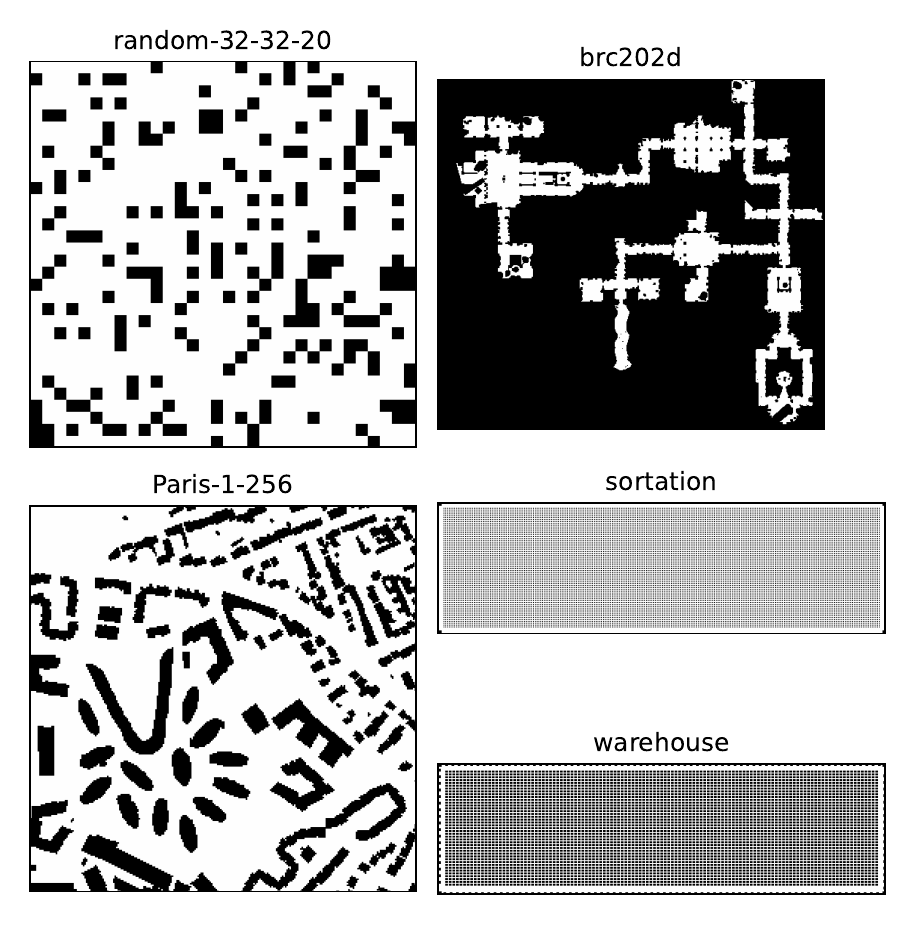}
	\caption{Visualization of maps used for the empirical evaluation.}
    \label{fig:maps}
\end{figure}

\begin{table}[t!]
    \centering
    \resizebox{\linewidth}{!}{
    \begin{tabular}{|c|c|c|c|c|c|}
        \hline
        Map & Size & $|V|$ & Agents num\\
        \hline
        random-32-32-20 & 32x32 & 819 & 100, 200, \dots, 800\\
        Paris-1-256 & 256x256 & 47240 & 1000, 2000, \dots, 10000\\
        brc202d & 481x530 & 43151 & 500, 1000, \dots, 5000\\
        sortation & 140x500 & 54320 & 1000, 2000, \dots, 10000\\
        warehouse & 140x500 & 38586 & 1000, 2000, \dots, 10000\\
        \hline
    \end{tabular}
    }
    \caption{Detailed information about evaluated maps and instances.}
    \label{tab:agents_num_info}
\end{table}
\begin{figure*}[t!]
	\includegraphics[width=\textwidth]{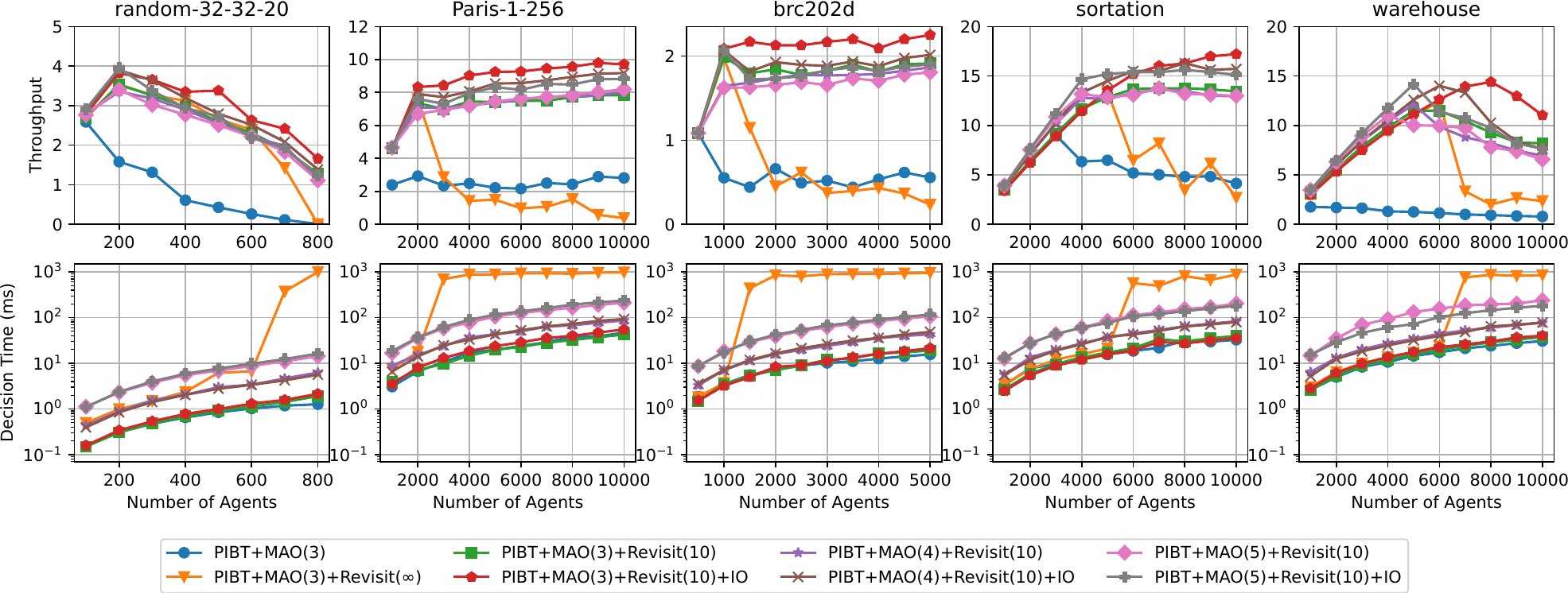}
	\caption{Ablation study of different enhancements incorporated into EPIBT and its evaluation with different operation lengths.}
    \label{fig:experiments_epibt}
\end{figure*}

\section{Empirical Evaluation}

We evaluated our approach on the same set of five maps used in the League of Robot Runners competition~\cite{chan2024league}: \texttt{random-32-32-20}, \texttt{brc202d}, \texttt{Paris-1-256}, \texttt{sortation}, and \texttt{warehouse}. Most maps originate from the well-known MAPF benchmark~\cite{stern2019multi}. Figure~\ref{fig:maps} shows their layouts. For each map, we ran multiple instances with varying agent counts, as detailed in Table~\ref{tab:agents_num_info}. Goal sequences and their order were taken directly from the competition's public archive.

To eliminate any inter-agent influence caused by goal assignment, we used the following logic: agent $k$ receives tasks with indices $k$, $k+\text{total\_agents}$, $k+2\cdot\text{total\_agents}$, etc., cycling through the task pool as needed. The time budget was set to 1 second per timestep for all experiments. All methods were run single-threaded on an Intel Xeon Gold 6338 CPU with 256~GB RAM. Timestep horizon $T$ for each experiment was set to 5{,}000 timesteps, except for \texttt{random-32-32-20}, which used 1{,}000.

\subsection{Ablation Study}
We first evaluated several EPIBT variants to assess the impact of each enhancement -- multi-action operations (MAO), revisiting, and inheritance of operations (IO) -- as well as the effect of operation length on runtime and performance. EPIBT was tested with operation lengths of 3, 4, and 5.

Results are shown in Figure~\ref{fig:experiments_epibt}. Here, MAO($x$) indicates operation length $x$, and Revisit($y$) denotes a revisit limit $L = y$. Adding MAO alone to PIBT yields poor performance, as agents cannot revise their choices. Introducing revisiting greatly improves results, but unlimited revisiting can cause the algorithm to hit the time budget, degrading performance. Limiting revisits stabilizes both throughput and runtime. Incorporating IO further boosts performance, regardless of operation length.

Longer operations increase runtime and only occasionally improve throughput (notably on large maps like \texttt{sortation} or \texttt{warehouse} with 5,000 agents). In most cases, length 3 yields the best throughput, likely because longer operations lead to more multi-agent collisions, which cannot be resolved by the current approach. Thus, agents may be forced to select suboptimal actions to avoid such collisions. Notably, even with 10,000 agents, EPIBT computes the next action in under 100~ms, with potential for further optimization.

In the next experiments, we use the fully enhanced version, denoted as EPIBT($x$), where $x$ is the operation length.

\subsection{EPIBT on LMAPF-T}

We compared EPIBT to PIBT-like approaches applicable to LMAPF-T. Since original PIBT cannot be directly used, we adapted it to consider only five operations -- FWW, RFW, CFW, RRF, and WWW -- mimicking the omnidirectional action model, where moving to some adjacent cells requires 2–3 actions. Thus, we do not consider MAO with length less than 3 on LMAPF-T, as shorter operations are insufficient. We also evaluated Causal PIBT, an event-based variant that handles multi-timestep actions. Aggregated throughput results are shown in Figure~\ref{fig:spider_plot}. This spider plot averages normalized results across all instances for each map and includes additional approaches discussed later. The results clearly show that EPIBT consistently outperforms both PIBT and Causal PIBT across all maps. %The detailed plots of this experiment are provided in the Appendix.

\begin{figure*}[t]
	\includegraphics[width=\textwidth]{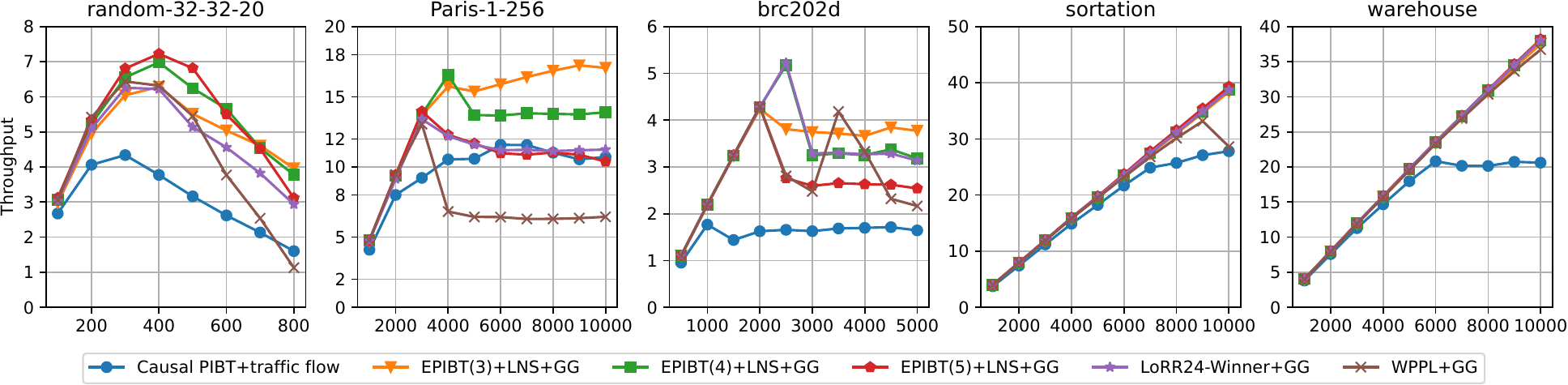}
	\caption{Comparison of EPIBT+LNS+GG with operation lengths 3, 4, and 5 with both winners of LoRR competition and Causal PIBT+traffic flow on the online LMAPF-T setting.}
    \label{fig:experiments_anytime_gg}
\end{figure*}

\begin{figure}[t!]
	\includegraphics[width=\linewidth]{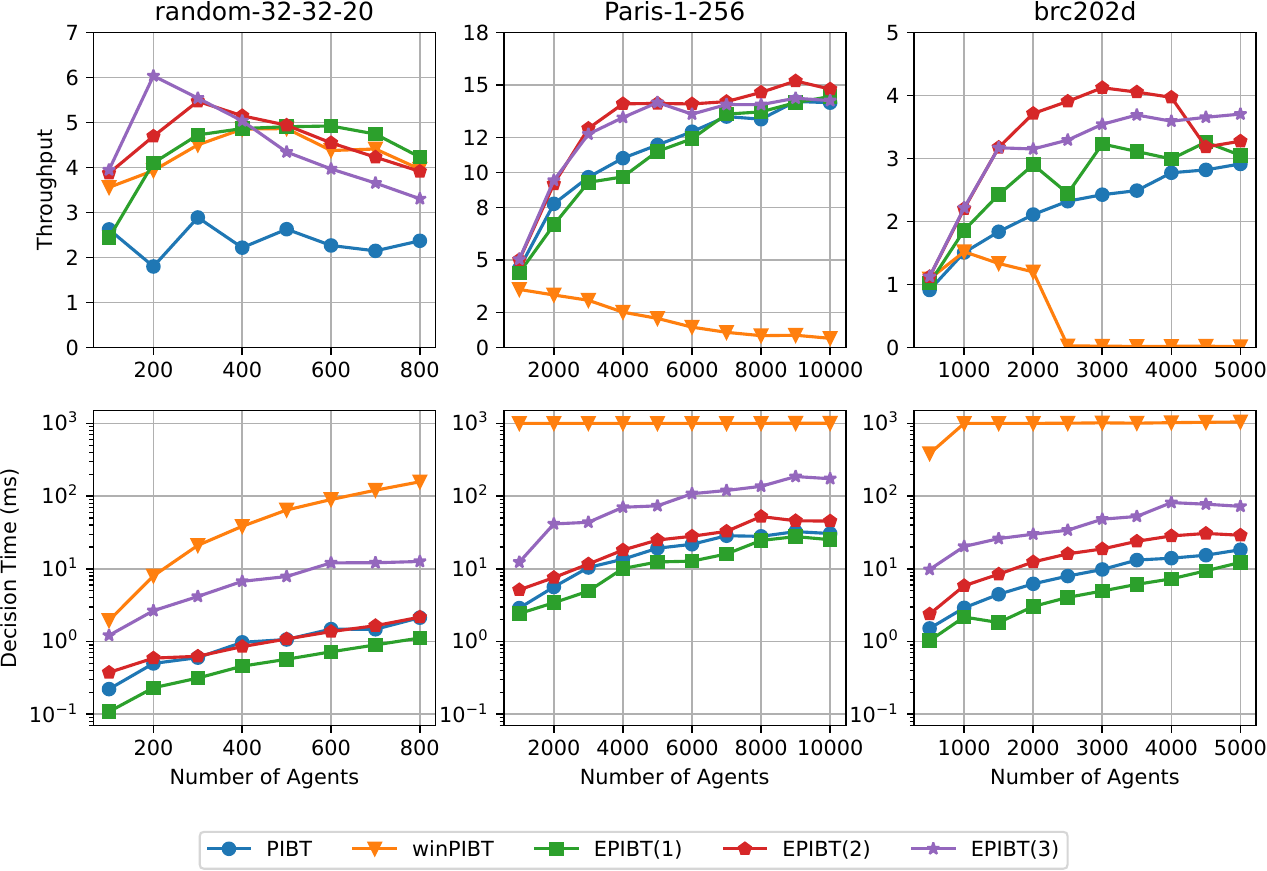}
	\caption{Comparison of EPIBT with original PIBT and winPIBT on the online LMAPF setting.}
    \label{fig:lmapf}
\end{figure}

\subsection{EPIBT on LMAPF}

We also evaluated EPIBT under the omnidirectional action model. Here, the number of unique cell sequences is $5^{\text{length}}$ and cannot be reduced, as each action leads to a distinct location. In this setting, EPIBT(1) and EPIBT(2) are feasible, since all adjacent cells are reachable in one step. As baselines, we tested original PIBT and winPIBT~\cite{okumura2019winpibt}, which plans for $w$ steps ahead. Following~\cite{okumura2019winpibt}, we set $w=10$, as this window size yields better results without runtime penalty.

Figure~\ref{fig:lmapf} presents the results. winPIBT performs well on the \texttt{random-32-32-20} map, but it reaches the time budget limit on the rest of the maps and as a result demonstrates poor performance. This likely occurs due to running A* for each agent, making it impractical for online LMAPF with a 1-second time budget. The original PIBT approach underperforms on \texttt{random-32-32-20}, but is able to compete with EPIBT(1) on \texttt{Paris-1-256}. This is explained by the fact that the \texttt{random-32-32-20} map contains deadlocks, i.e., some of the cells have no cycle of 3+ size. As a result, some of the agents fail quite frequently and PIBT cannot rechoose their actions, as revisiting is not allowed. Such deadlocks are not present on large maps, which partially eliminates the advantage that EPIBT(1) gains from revisits. Utilizing operation length 2 for LMAPF is definitely beneficial. However, further increments in operation length result in a significant increase in runtime and do not always result in improvement in terms of throughput. Due to space limitations, the results on only 3 out of 5 maps are displayed. However, the results on the remaining 2 maps are similar to the ones obtained on \texttt{Paris-1-256} map.

\subsection{EPIBT+LNS}

We finally compared EPIBT combined with LNS to state-of-the-art approaches for online LMAPF-T. The first baseline is WPPL~\cite{jiang2024scaling}, the winning solution of LoRR23. The second is Causal PIBT with traffic flow~\cite{chen2024traffic}, used as the default planner in LoRR24. The third, which we refer to as LoRR24-Winner\footnote{Implementation was taken from https://github.com/MAPF-Competition/Code-Archive}, is the winning solution of LoRR24. The latter is a method created by us for the competition. It is also based on an EPIBT variant, but it has no operation inheritance and limits agent revisits in a different way. Both WPPL and LoRR24-Winner employ Graph Guidance (GG), which reduces congestion by weighting transition costs. We also incorporated GG into EPIBT+LNS, using the same weights as LoRR24-Winner. All methods were evaluated with and without GG.

Aggregated results are shown in Figure~\ref{fig:spider_plot}. EPIBT(3)+LNS+GG achieves score 1.0 on 3 out of 5 maps, demonstrating that none of the competitors is able to outperform it in those maps. On the remaining two maps (random-32-32-20 and brc202d), other methods occasionally achieve higher scores, particularly on low-agent instances. However, EPIBT(3)+LNS+GG still outperforms others in most cases. Detailed results, including EPIBT+LNS+GG with operation lengths 4 and 5, are provided in Figure \ref{fig:experiments_anytime_gg}. %The detailed results without GG are provided in the Appendix. 
No runtime plot is shown, as all approaches used the full 1-second time budget per timestep.

Overall, these results demonstrate that EPIBT, when combined with LNS and GG, can outperform current state-of-the-art methods for online LMAPF-T.

\section{Conclusion}

We introduced EPIBT, a new modification of the PIBT approach tailored for the rotation action model. The key innovation is the use of multi-action operations, enabling the algorithm to consider multiple timesteps at once without invoking a separate path-planning routine, unlike windowed methods such as winPIBT. Additional enhancements, including limited revisiting and operation inheritance, further improve performance and efficiency.

Empirical results show that these enhancements significantly boost throughput while maintaining fast runtime. Moreover, combining EPIBT with Large Neighborhood Search and Graph Guidance yields performance surpassing all existing methods, including both League of Robots Runners competition winners, establishing a new state-of-the-art for online LMAPF-T. Future work includes improving multi-agent collision handling and further optimizing the integration with LNS and GG.

\bibliography{aaai2026}

@inproceedings{li2021lifelong,
  title={Lifelong multi-agent path finding in large-scale warehouses},
  author={Li, Jiaoyang and Tinka, Andrew and Kiesel, Scott and Durham, Joseph W and Kumar, TK Satish and Koenig, Sven},
  booktitle={Proceedings of the AAAI Conference on Artificial Intelligence},
  volume={35},
  pages={11272--11281},
  year={2021}
}

@inproceedings{silver2005cooperative,
  title={Cooperative pathfinding},
  author={Silver, David},
  booktitle={Proceedings of the aaai conference on artificial intelligence and interactive digital entertainment},
  volume={1},
  pages={117--122},
  year={2005}
}

@inproceedings{jiang2024scaling,
  title={Scaling lifelong multi-agent path finding to more realistic settings: Research challenges and opportunities},
  author={Jiang, He and Zhang, Yulun and Veerapaneni, Rishi and Li, Jiaoyang},
  booktitle={Proceedings of the International Symposium on Combinatorial Search},
  volume={17},
  pages={234--242},
  year={2024}
}

@inproceedings{li2021anytime,
  title={Anytime multi-agent path finding via large neighborhood search},
  author={Li, Jiaoyang and Chen, Zhe and Harabor, Daniel and Stuckey, Peter J and Koenig, Sven},
  booktitle={International Joint Conference on Artificial Intelligence 2021},
  pages={4127--4135},
  year={2021}
}

@article{sharon2015conflict,
  title={Conflict-based search for optimal multi-agent pathfinding},
  author={Sharon, Guni and Stern, Roni and Felner, Ariel and Sturtevant, Nathan R},
  journal={Artificial intelligence},
  volume={219},
  pages={40--66},
  year={2015},
  publisher={Elsevier}
}

@inproceedings{boyarski2015icbs,
  title={Icbs: The improved conflict-based search algorithm for multi-agent pathfinding},
  author={Boyarski, Eli and Felner, Ariel and Stern, Roni and Sharon, Guni and Betzalel, Oded and Tolpin, David and Shimony, Eyal},
  booktitle={Proceedings of the International Symposium on Combinatorial Search},
  volume={6},
  pages={223--225},
  year={2015}
}

@inproceedings{li2019improved,
  title={Improved Heuristics for Multi-Agent Path Finding with Conflict-Based Search.},
  author={Li, Jiaoyang and Felner, Ariel and Boyarski, Eli and Ma, Hang and Koenig, Sven},
  booktitle={IJCAI},
  volume={2019},
  pages={442--449},
  year={2019}
}

@inproceedings{huang2021learning,
  title={Learning node-selection strategies in bounded suboptimal conflict-based search for multi-agent path finding},
  author={Huang, Taoan and Dilkina, Bistra and Koenig, Sven},
  booktitle={International joint conference on autonomous agents and multiagent systems (AAMAS)},
  year={2021}
}

@inproceedings{barer2014suboptimal,
  title={Suboptimal variants of the conflict-based search algorithm for the multi-agent pathfinding problem},
  author={Barer, Max and Sharon, Guni and Stern, Roni and Felner, Ariel},
  booktitle={Proceedings of the international symposium on combinatorial Search},
  volume={5},
  pages={19--27},
  year={2014}
}

@article{sharon2013increasing,
  title={The increasing cost tree search for optimal multi-agent pathfinding},
  author={Sharon, Guni and Stern, Roni and Goldenberg, Meir and Felner, Ariel},
  journal={Artificial intelligence},
  volume={195},
  pages={470--495},
  year={2013},
  publisher={Elsevier}
}

@article{lam2022branch,
  title={Branch-and-cut-and-price for multi-agent path finding},
  author={Lam, Edward and Le Bodic, Pierre and Harabor, Daniel and Stuckey, Peter J},
  journal={Computers \& Operations Research},
  volume={144},
  pages={105809},
  year={2022},
  publisher={Elsevier}
}

@inproceedings{li2022mapf,
  title={MAPF-LNS2: Fast repairing for multi-agent path finding via large neighborhood search},
  author={Li, Jiaoyang and Chen, Zhe and Harabor, Daniel and Stuckey, Peter J and Koenig, Sven},
  booktitle={Proceedings of the AAAI Conference on Artificial Intelligence},
  volume={36},
  pages={10256--10265},
  year={2022}
}

@inproceedings{li2021eecbs,
  title={Eecbs: A bounded-suboptimal search for multi-agent path finding},
  author={Li, Jiaoyang and Ruml, Wheeler and Koenig, Sven},
  booktitle={Proceedings of the AAAI conference on artificial intelligence},
  volume={35},
  pages={12353--12362},
  year={2021}
}

@inproceedings{huang2022anytime,
  title={Anytime multi-agent path finding via machine learning-guided large neighborhood search},
  author={Huang, Taoan and Li, Jiaoyang and Koenig, Sven and Dilkina, Bistra},
  booktitle={Proceedings of the AAAI Conference on Artificial Intelligence},
  volume={36},
  pages={9368--9376},
  year={2022}
}

@inproceedings{okumura2023lacam,
  title={Lacam: Search-based algorithm for quick multi-agent pathfinding},
  author={Okumura, Keisuke},
  booktitle={Proceedings of the AAAI Conference on Artificial Intelligence},
  volume={37},
  pages={11655--11662},
  year={2023}
}

@article{de2014push,
  title={Push and rotate: a complete multi-agent pathfinding algorithm},
  author={De Wilde, Boris and Ter Mors, Adriaan W and Witteveen, Cees},
  journal={Journal of Artificial Intelligence Research},
  volume={51},
  pages={443--492},
  year={2014}
}

@inproceedings{okumura2024engineering,
  title={Engineering LaCAM*: Towards Real-time, Large-scale, and Near-optimal Multi-agent Pathfinding},
  author={Okumura, Keisuke},
  booktitle={Proceedings of the 23rd International Conference on Autonomous Agents and Multiagent Systems},
  pages={1501--1509},
  year={2024}
}

@article{okumura2022priority,
  title={Priority inheritance with backtracking for iterative multi-agent path finding},
  author={Okumura, Keisuke and Machida, Manao and D{\'e}fago, Xavier and Tamura, Yasumasa},
  journal={Artificial Intelligence},
  volume={310},
  pages={103752},
  year={2022},
  publisher={Elsevier}
}

@inproceedings{stern2019multi,
  title={Multi-agent pathfinding: Definitions, variants, and benchmarks},
  author={Stern, Roni and Sturtevant, Nathan and Felner, Ariel and Koenig, Sven and Ma, Hang and Walker, Thayne and Li, Jiaoyang and Atzmon, Dor and Cohen, Liron and Kumar, TK and others},
  booktitle={Proceedings of the International Symposium on Combinatorial Search},
  volume={10},
  pages={151--158},
  year={2019}
}

@inproceedings{chen2024traffic,
  title={Traffic flow optimisation for lifelong multi-agent path finding},
  author={Chen, Zhe and Harabor, Daniel and Li, Jiaoyang and Stuckey, Peter J},
  booktitle={Proceedings of the AAAI Conference on Artificial Intelligence},
  volume={38},
  pages={20674--20682},
  year={2024}
}

@inproceedings{chan2024league,
  title={The League of Robot Runners Competition: Goals, Designs, and Implementation},
  author={Chan, Shao-Hung and Chen, Zhe and Guo, Teng and Zhang, Han and Zhang, Yue and Harabor, Daniel and Koenig, Sven and Wu, Cathy and Yu, Jingjin},
  booktitle={ICAPS 2024 System's Demonstration track},
  year={2024}
}

@inproceedings{geft2022refined,
  title={Refined Hardness of Distance-Optimal Multi-Agent Path Finding},
  author={Geft, Tzvika and Halperin, Dan},
  booktitle={Proceedings of the 21st International Conference on Autonomous Agents and Multiagent Systems},
  pages={481--488},
  year={2022}
}

@incollection{dhaliwal2020rise,
  title={The rise of automation and robotics in warehouse management},
  author={Dhaliwal, Amandeep},
  booktitle={Transforming management using artificial intelligence techniques},
  pages={63--72},
  year={2020},
  publisher={CRC Press}
}

@article{ferreira2023systematic,
  title={A systematic literature review on the application of automation in logistics},
  author={Ferreira, B{\'a}rbara and Reis, Jo{\~a}o},
  journal={Logistics},
  volume={7},
  pages={80},
  year={2023},
  publisher={MDPI}
}

@inproceedings{okumura2021time,
  title={Time-independent planning for multiple moving agents},
  author={Okumura, Keisuke and Tamura, Yasumasa and D{\'e}fago, Xavier},
  booktitle={Proceedings of the AAAI Conference on Artificial Intelligence},
  volume={35},
  pages={11299--11307},
  year={2021}
}

@article{okumura2019winpibt,
  title={winpibt: Extended prioritized algorithm for iterative multi-agent path finding},
  author={Okumura, Keisuke and Tamura, Yasumasa and D{\'e}fago, Xavier},
  journal={arXiv preprint arXiv:1905.10149},
  year={2019}
}

@inproceedings{veerapaneni2025windowed,
  title={Windowed MAPF with Completeness Guarantees},
  author={Veerapaneni, Rishi and Saleem, Muhammad Suhail and Li, Jiaoyang and Likhachev, Maxim},
  booktitle={Proceedings of the AAAI Conference on Artificial Intelligence},
  volume={39},
  pages={23323--23332},
  year={2025}
}
\clearpage
\appendix
\appendix

\section*{Appendix}

\section{Extra Empirical Evaluation}

\subsection{Detailed Plots}
\begin{figure*}[t]
	\includegraphics[width=\textwidth]{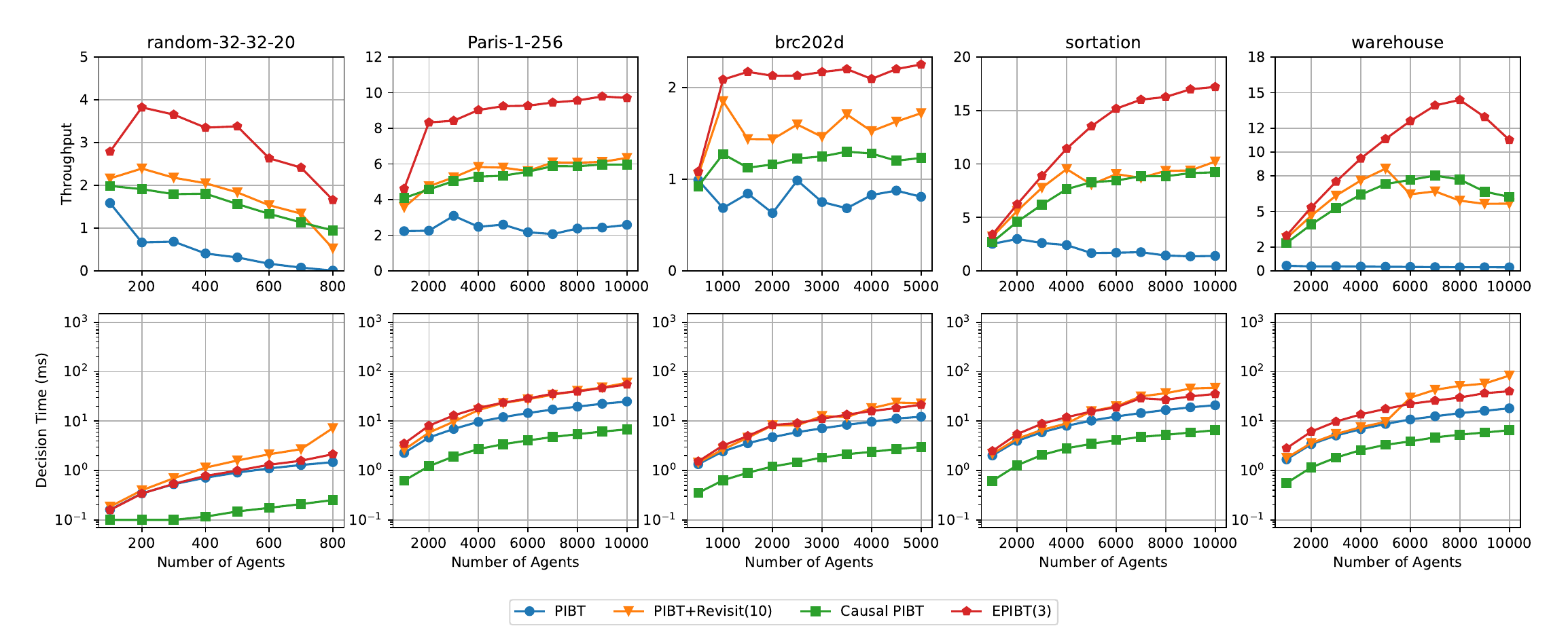}
	\caption{The results of evaluation of EPIBT(3) compared to other PIBT-like approaches on online LMAPF-T setting.}
    \label{fig:experiments_pibt}
\end{figure*}

Figure~\ref{fig:experiments_pibt} presents the detailed results of the experiment where EPIBT was compared to other PIBT-like approaches on online LMAPF-T. EPIBT(3) was compared against PIBT, Causal PIBT, and PIBT with allowed revisit. Across all maps, regardless of the number of agents in the instances, EPIBT(3) significantly outperforms all baselines. In terms of runtime, Causal PIBT achieves better results. This difference is explained by different implementations: we utilized a publicly available implementation of Causal PIBT that is heavily optimized and specifically tailored to solve LMAPF-T. In contrast, our EPIBT implementation is designed to support both LMAPF and LMAPF-T, different operation lengths, and other features. In other words, the EPIBT implementation is more versatile but consequently less computationally efficient.

\begin{figure*}[t]
	\includegraphics[width=\textwidth]{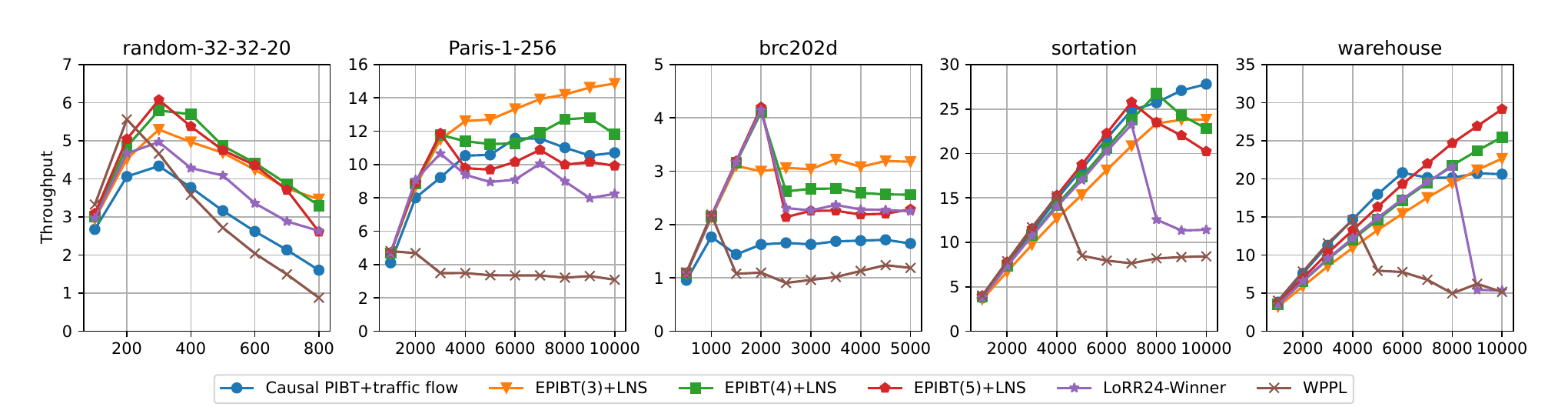}
	\caption{The results of evaluation of EPIBT+LNS compared to baselines that also utilize LNS.}
    \label{fig:experiments_anytime}
\end{figure*}

Figure~\ref{fig:experiments_anytime} presents the detailed results of the experiment where EPIBT+LNS with different operation lengths was compared against WPPL, LoRR24-Winner, and Causal PIBT+traffic flow. Although the latter utilizes dynamic graph guidance instead of an LNS optimizer, its results are included in this figure as well. In most cases, EPIBT+LNS outperforms the evaluated baselines regardless of the operation length used. Similar to when EPIBT is additionally combined with GG, the most effective operation length varies depending on the map used. On \texttt{Paris-1-256} and \texttt{brc202d} maps, the best-performing approach is EPIBT(3)+LNS. This occurs due to high congestion on these maps, where crowds of agents accumulate in narrow passages in the middle of the map. Visualizations of such crowded areas are shown in the next section.

\subsection{Tie-breaking Ablation Study}
EPIBT requires a tie-breaking mechanism for operations with equal goal distance. We evaluated all operation lengths with different tie-breaking strategies using notation EPIBT(3)+FRW, where F (forward) is most preferred, followed by R (rotation), then W (wait). We also tested random tie-breaking (RND) and no tie-breaking (-OPW).

\begin{figure*}[t]
	\includegraphics[width=\textwidth]{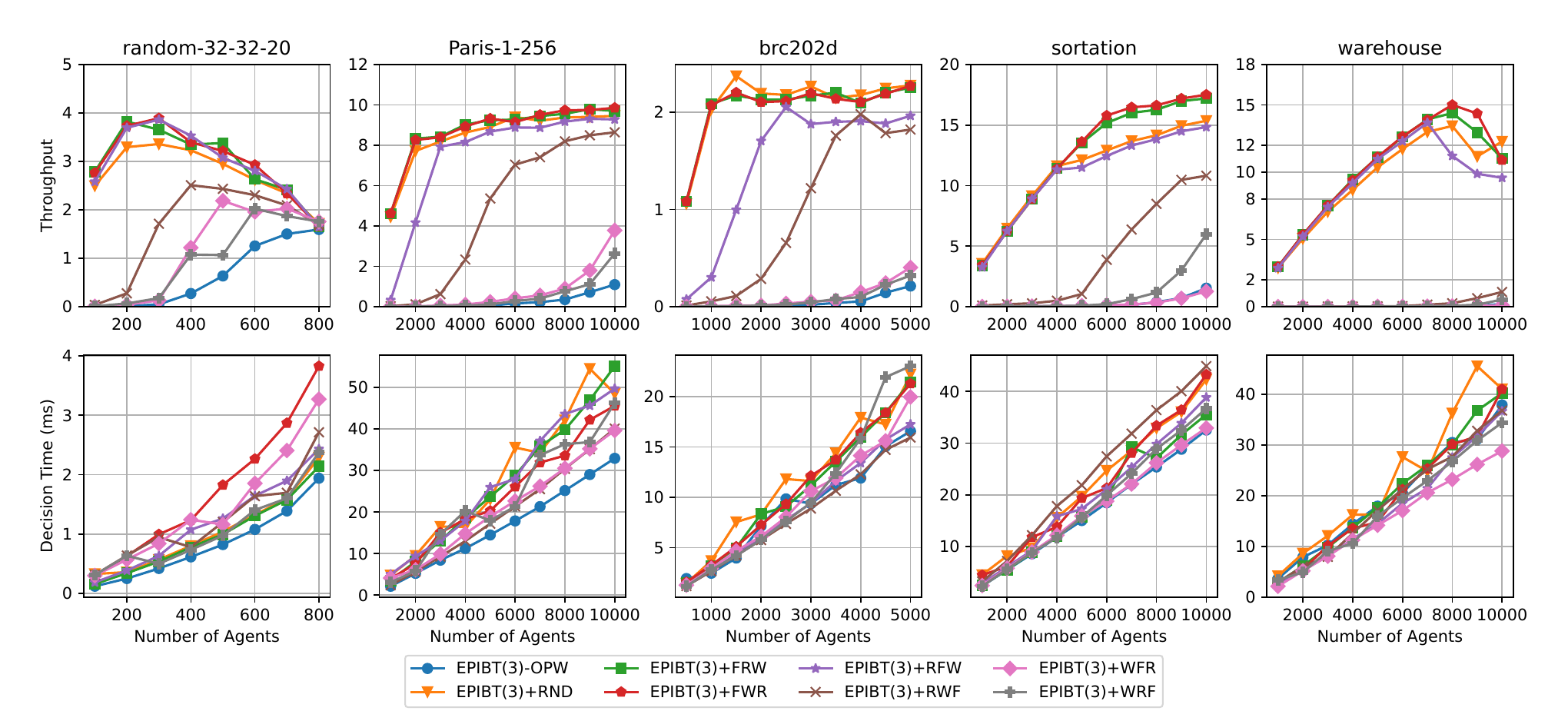}
	\caption{The results of evaluation of EPIBT(3) with various tie-breaking mechanisms utilized during operations sorting.}
    \label{fig:metric_plot_opw_3}
\end{figure*}

\begin{figure*}[t]
	\includegraphics[width=\textwidth]{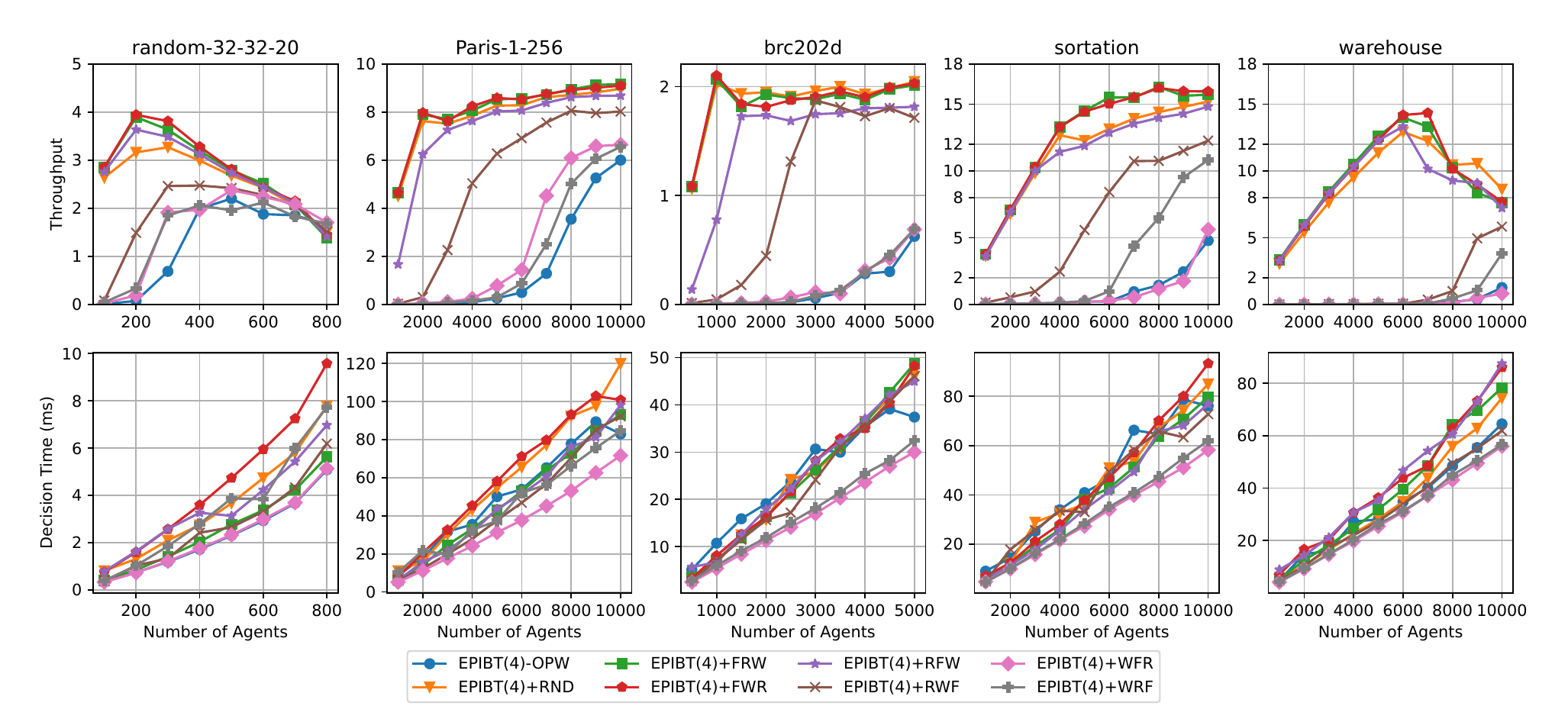}
	\caption{The results of evaluation of EPIBT(4) with various tie-breaking mechanisms utilized during operations sorting.}
    \label{fig:metric_plot_opw_4}
\end{figure*}

\begin{figure*}[t]
	\includegraphics[width=\textwidth]{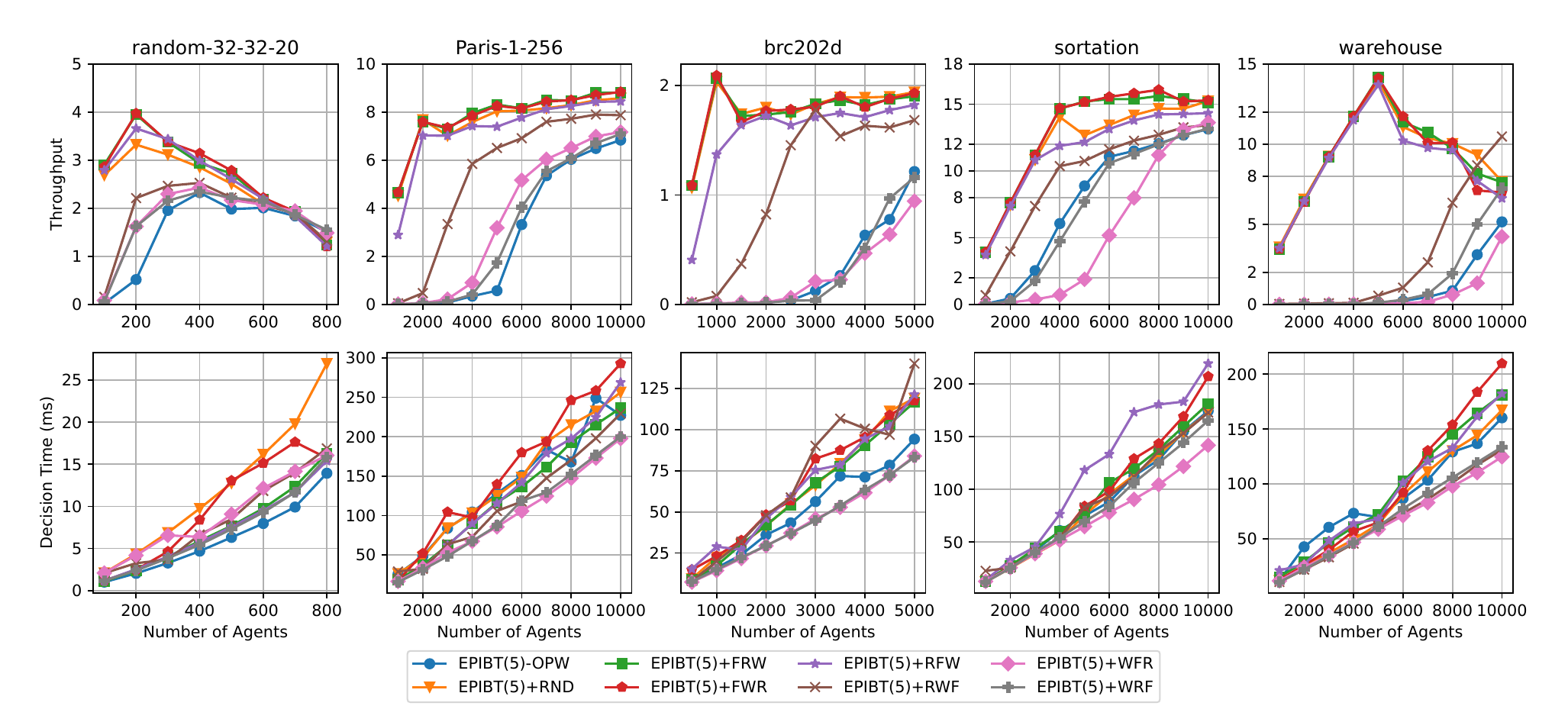}
	\caption{The results of evaluation of EPIBT(5) with various tie-breaking mechanisms utilized during operations sorting.}
    \label{fig:metric_plot_opw_5}
\end{figure*}

The results of this experiment for EPIBT(3) are shown in Figure~\ref{fig:metric_plot_opw_3}, for EPIBT(4) in Figure~\ref{fig:metric_plot_opw_4}, and for EPIBT(5) in Figure~\ref{fig:metric_plot_opw_5}. Regardless of operation length, there are three best-performing tie-breaking mechanisms: FRW, FWR, and surprisingly, RND. This experiment demonstrates that it is crucial to prefer move actions over rotate and wait actions, while the preference between wait and rotation actions is negligible. The good performance of RND is explained by its stochastic nature, which helps resolve deadlocks and failure cases, as failed agents may have different operation preference orders at each timestep. The remaining evaluated tie-breaking mechanisms, where either wait or rotate actions are more preferable than move forward actions, significantly underperform, especially on instances with few agents. The case without any tie-breaking mechanism also demonstrates poor results because, unlike RND, it lacks stochasticity and agents simply have bad operation ordering.

The poor performance of tie-breaking variants that underprefer move actions is explained by the fact that it is important to force agents to move. For example, operations WFW, WWF, and FWW all yield the same distance to goal. However, if an agent chooses WFW or WWF instead of FWW, it will remain stationary, and on the next step it may again select an action that begins with a wait action. As a result, the agent becomes stuck even in the absence of other agents. This example is not artificial, it may occur when the goal location is in an adjacent cell and the agent needs to perform only a single forward move action.

\subsection{Revisit Ablation Study}
EPIBT also includes a parameterized enhancement called revisit, which allows agents to be visited multiple times to consider different actions. This is necessary because different operations may result in collisions with the same agents and may help improve cooperation between agents. However, infinite revisit dramatically increases the algorithm's runtime. Therefore, we introduced a limit $L$ and used this value in all subsequent experiments. We have already demonstrated the difference between infinite and limited revisit in the first series of experiments presented in the main text. Here, we present results for different limits. We evaluated various revisit limits with all three operation lengths: EPIBT(3), EPIBT(4), and EPIBT(5).

\begin{figure*}[t]
	\includegraphics[width=\textwidth]{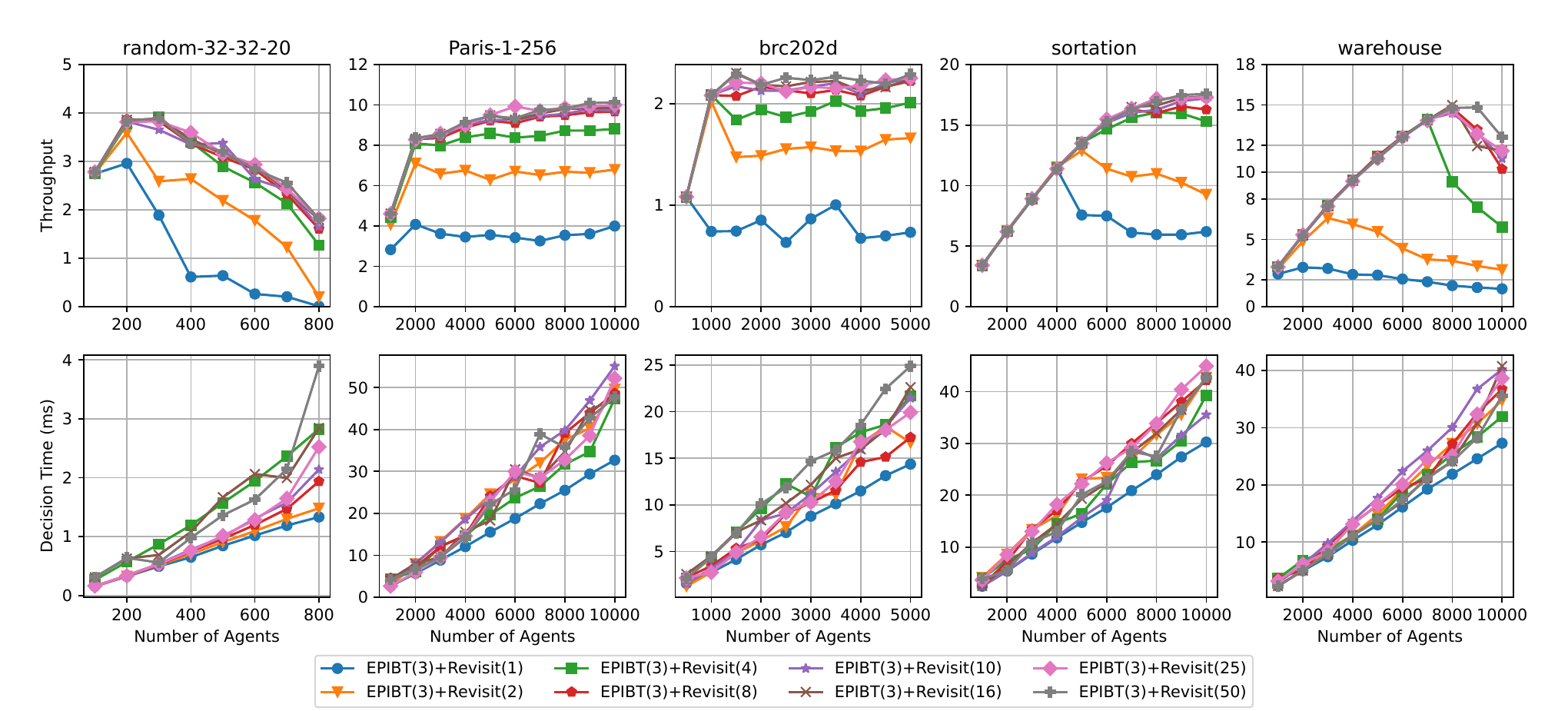}
	\caption{The results of EPIBT(3) with various revisit limits.}
    \label{fig:experiments_epibt_revisit_3}
\end{figure*}

\begin{figure*}[t]
	\includegraphics[width=\textwidth]{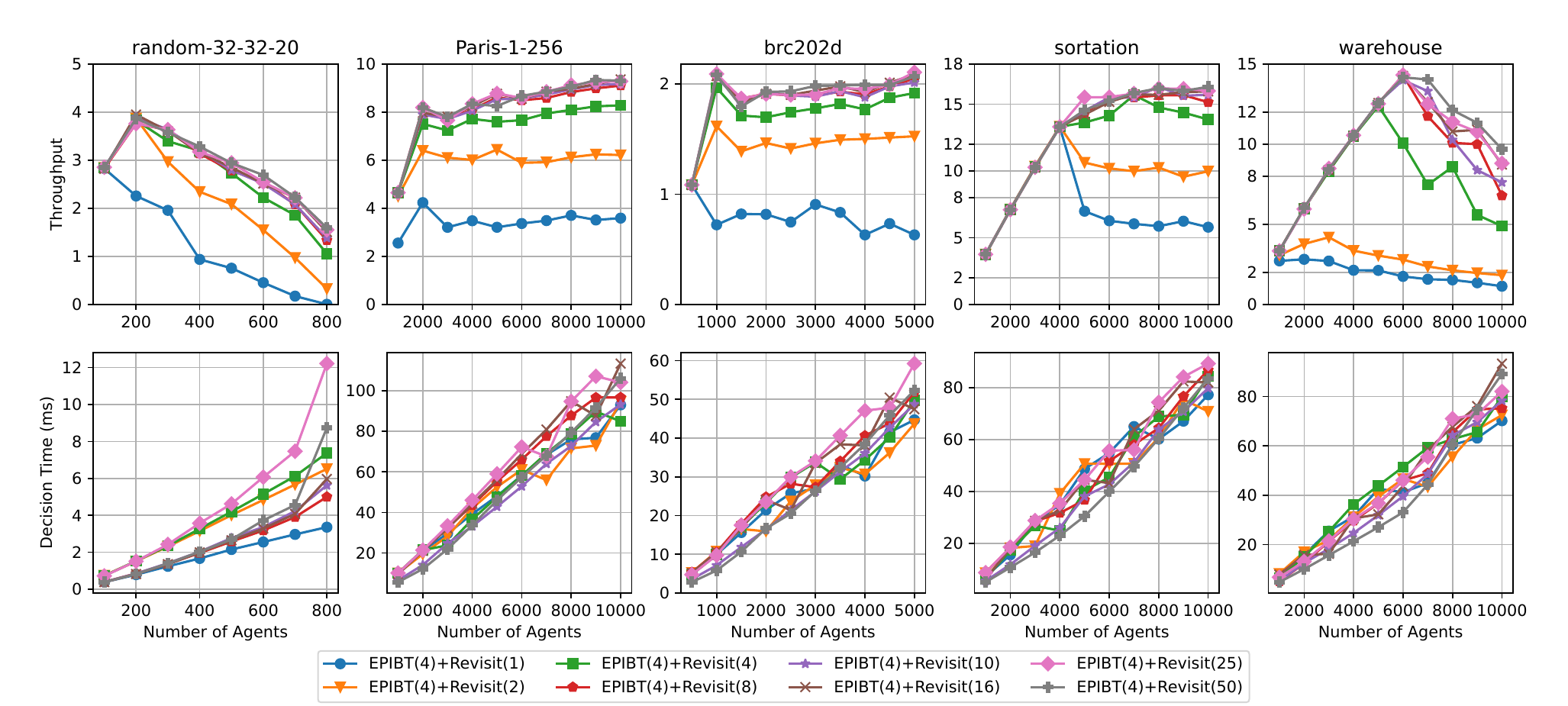}
	\caption{The results of EPIBT(4) with various revisit limits.}
    \label{fig:experiments_epibt_revisit_4}
\end{figure*}

\begin{figure*}[t]
	\includegraphics[width=\textwidth]{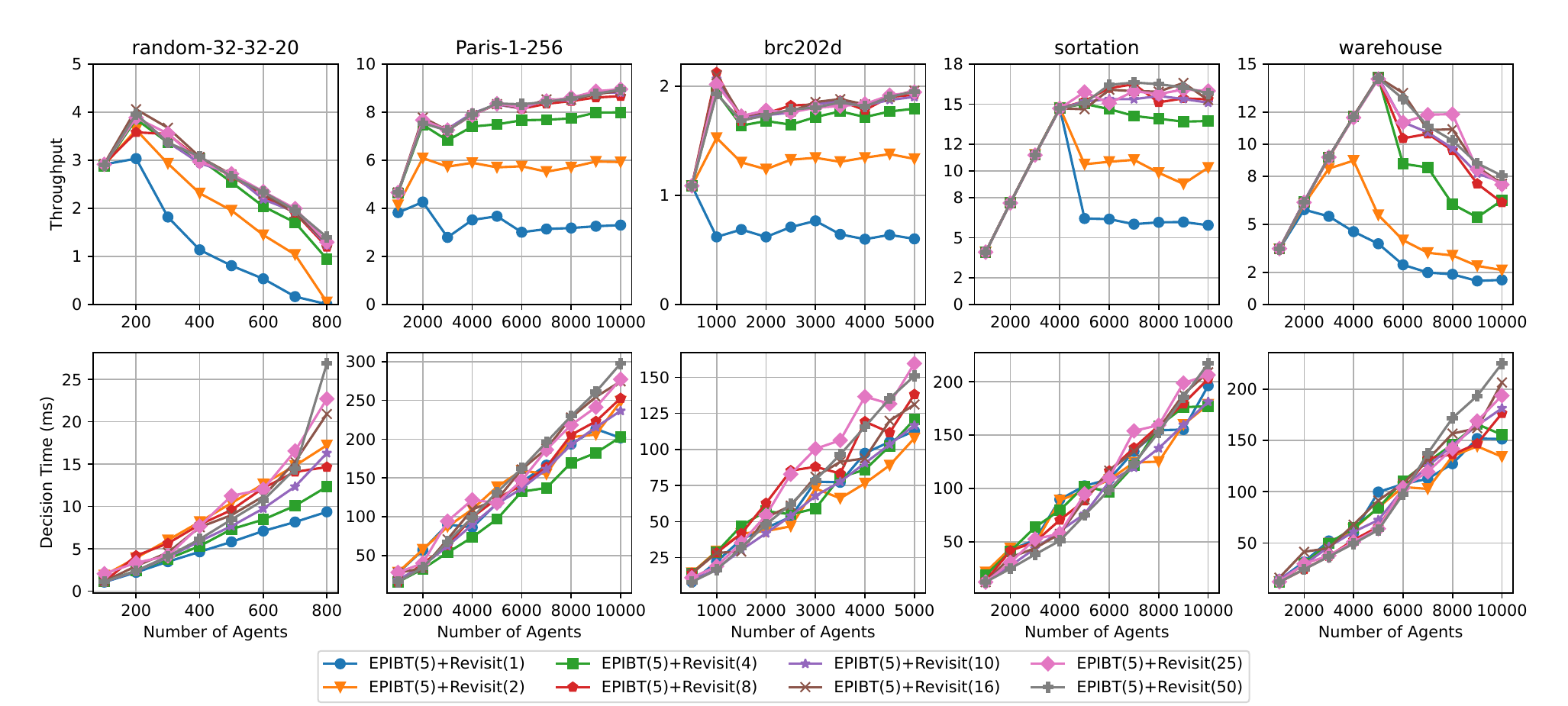}
	\caption{The results of EPIBT(5) with various revisit limits.}
    \label{fig:experiments_epibt_revisit_5}
\end{figure*}

The results of this experiment are depicted in Figures~\ref{fig:experiments_epibt_revisit_3}, \ref{fig:experiments_epibt_revisit_4}, and \ref{fig:experiments_epibt_revisit_5}, respectively. Regardless of operation length, it is clearly seen that only limits 1, 2, and 4 underperform, while the remaining evaluated limits (8, 10, 16, 25, and 50) demonstrate relatively similar results. This experiment clearly demonstrates that high revisit limits do not lead to throughput improvement. One simply needs to utilize a sufficiently high revisit limit, and our chosen value of 10 is adequate.

\subsection{Waiting Heatmaps}

In this section, we graphically demonstrate the benefit of utilizing the Graph Guidance technique. The heatmaps visualize waiting actions: the more red a cell appears, the more timesteps agents spend waiting in that location. The presence of yellow and especially red regions indicates that agents frequently get stuck in these areas, which is detrimental to overall throughput. In contrast, areas where agents move through efficiently (without excessive waiting) appear in green color.

\begin{figure*}[t]
	\includegraphics[width=\textwidth]{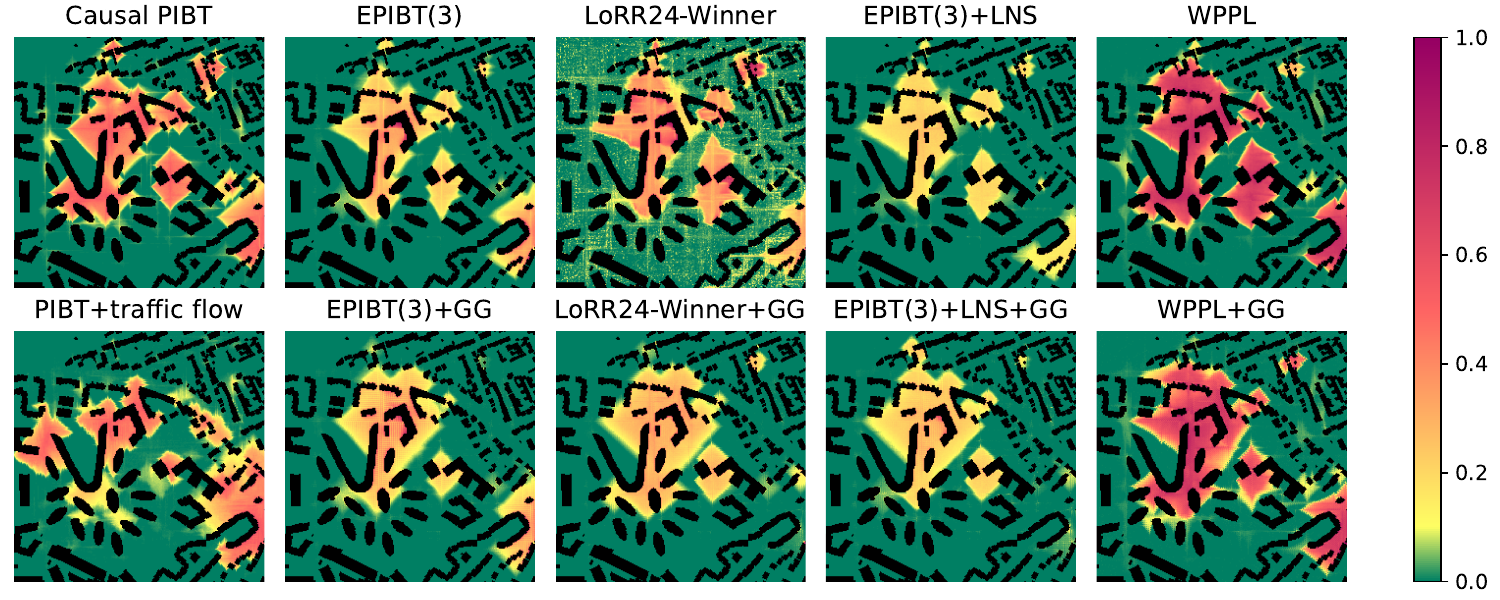}
	\caption{Waiting heatmaps of various approaches for \texttt{Paris-1-256} map and an instance with 10,000 agents.}
    \label{fig:heatmap_city}
\end{figure*}

\begin{figure*}[t]
	\includegraphics[width=\textwidth]{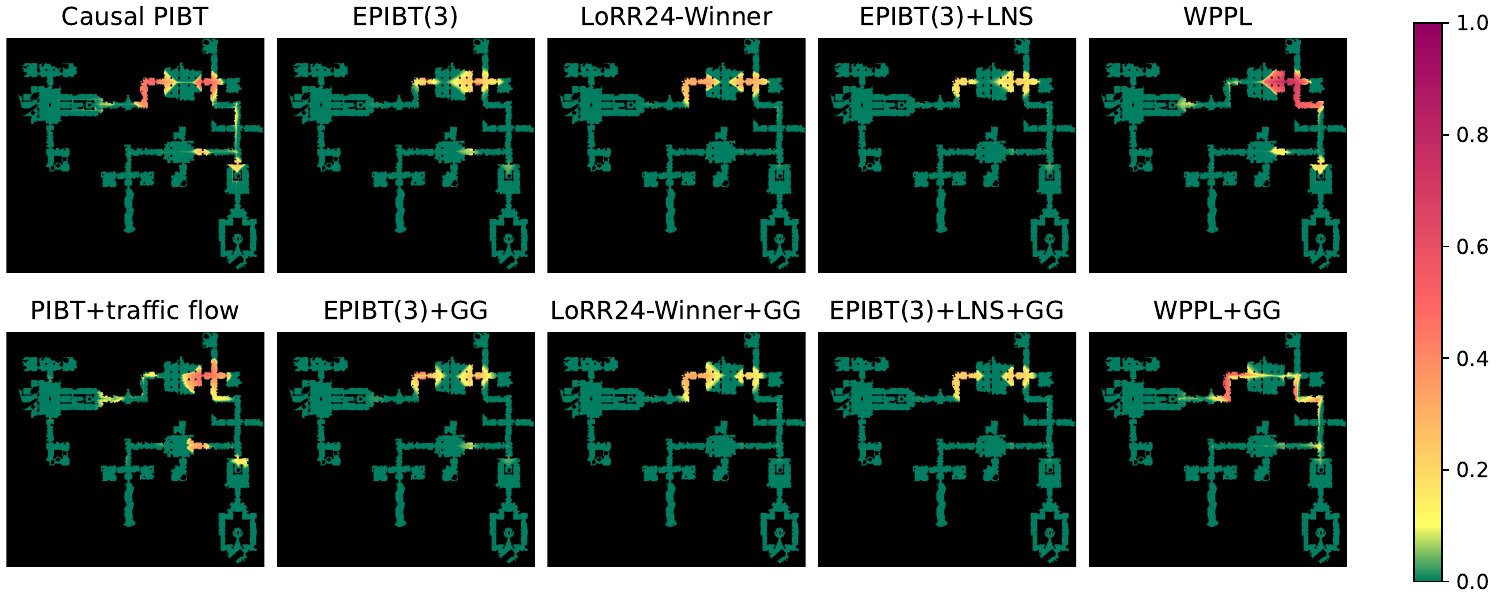}
	\caption{Waiting heatmaps of various approaches for \texttt{brc202d} map and an instance with 3500 agents.}
    \label{fig:heatmap_game}
\end{figure*}

\begin{figure*}[t]
	\includegraphics[width=\textwidth]{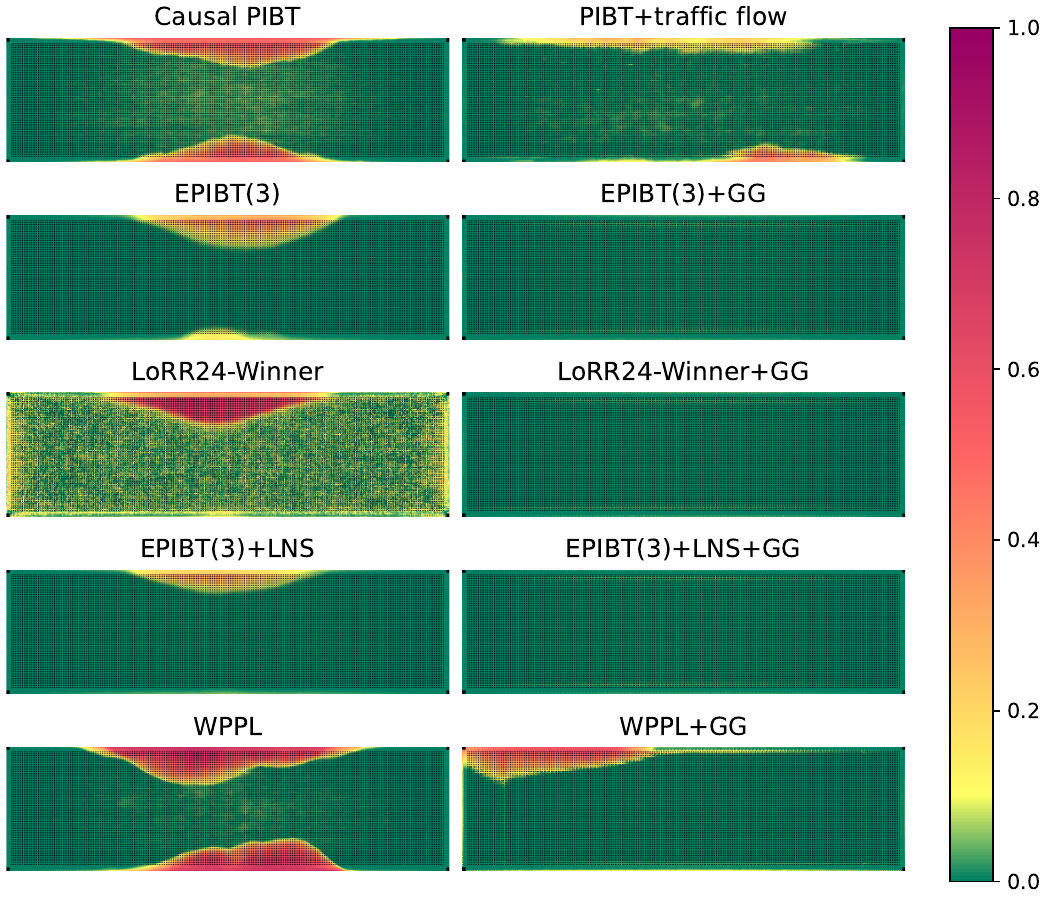}
	\caption{Waiting heatmaps of various approaches for \texttt{sortation} map and an instance with 10,000 agents.}
    \label{fig:heatmap_sortation}
\end{figure*}

Figure~\ref{fig:heatmap_city} demonstrates the heatmap of various approaches on the \texttt{Paris-1-256} map, Figure~\ref{fig:heatmap_game} shows results on the \texttt{brc202d} map, and Figure~\ref{fig:heatmap_sortation} presents results on the \texttt{sortation} map. Across all maps, EPIBT(3)+LNS exhibits the least amount of yellow/red regions among all evaluated approaches that do not utilize GG or traffic flow. This indicates that even without GG, EPIBT is able to efficiently resolve collisions and prevent agents from getting stuck in highly congested areas. Adding GG further helps approaches reduce agent waiting times and distribute traffic more evenly across the map. As a result, the combination of EPIBT+LNS with GG achieves the most efficient agent flow with minimal waiting-induced bottlenecks.

%\subsection{Source Code}
%The source code of the proposed approach is provided in the supplementary material accompanying this submission. All instances and maps used in the experimental evaluation are also included. We will make the code publicly available to enable other researchers to verify and reproduce our results, and to facilitate the development of enhanced planners building upon the EPIBT framework.

\end{document}